\documentclass[preprint,12pt]{elsarticle}

\usepackage{amsmath,amsfonts,amssymb}
\usepackage{algorithm,algorithmic}
\usepackage{graphicx,subfig,stfloats,booktabs,diagbox,multirow,url,hyperref}
\usepackage{cleveref}

\begin{document}

\begin{frontmatter}

\title{M-ary Precomputation-Based Accelerated Scalar Multiplication Algorithms for Enhanced Elliptic Curve Cryptography}


\author[1]{Tongxi Wu\corref{equal1}}  
\author[1]{Xufeng Liu\corref{equal1}}
\author[1]{Jin Yang\corref{cor1}}  
\author[2]{Yijie Zhu}
\author[1]{Shunyang Zeng}
\author[1]{Mingming Zhan}

\cortext[equal1]{These authors contributed equally to this work.}
\cortext[cor1]{Corresponding author. Email: wtx2021141230137@stu.scu.edu.cn}

\address[1]{College of Cyber Science and Engineering, Sichuan University, Chengdu 610207, China}
\address[2]{Institute of Software, Chinese Academy of Sciences and University of Chinese Academy of Sciences, Beijing 100190, China}

\begin{abstract}
Efficient scalar multiplication is critical for enhancing the performance of elliptic curve cryptography (ECC), especially in applications requiring large-scale or real-time cryptographic operations. This paper proposes an M-ary precomputation-based scalar multiplication algorithm, aiming to optimize both computational efficiency and memory usage. The method reduces the time complexity from $\Theta(Q \log p)$ to $\Theta\left(\frac{Q \log p}{\log Q}\right)$ and achieves a memory complexity of $\Theta\left(\frac{Q \log p}{\log^2 Q}\right)$. Experiments on ElGamal encryption and NS3-based communication simulations validate its effectiveness. On secp256k1, the proposed method achieves up to a 59\% reduction in encryption time and 30\% memory savings. In network simulations, the binary-optimized variant reduces communication time by 22.1\% on secp384r1 and simulation time by 25.4\% on secp521r1. The results demonstrate the scalability, efficiency, and practical applicability of the proposed algorithm. The source code will be publicly released upon acceptance.
\end{abstract}

\begin{keyword}
Scalar Multiplication \sep ECC \sep Cryptology \sep Precomputation
\end{keyword}

\end{frontmatter}

\section{Introduction}
Information technology has become one of the fundamental pillars of modern society, enabling innovations such as distributed computing, databases, and blockchain systems~\cite{das2024asynchronous,dolev2023sodsbc}. As the value of information systems increases, they have become prominent targets for cyberattacks. Cryptographic techniques are essential to protect these systems, ensuring confidentiality, integrity, and availability. In particular, asymmetric cryptographic algorithms are widely adopted due to their advantages in key management and security strength~\cite{bandarupalli2024random}. Furthermore, they provide critical digital signature capabilities~\cite{duan2023fin}, which are foundational to applications such as cryptocurrencies and decentralized finance.

Elliptic curve cryptography (ECC) ~\cite{cao2024practical} is widely employed in information system encryption because of its robust security, high efficiency, and compact key size ~\cite{dahiphale2025securing}. ECC has garnered attention for its effectiveness in various applications, including financial transactions, network security, and data protection\cite{ullah2023elliptic}. In these applications, enhancing the efficiency of ECC is crucial, particularly in scenarios that require quick responses and high security. A key aspect of enhancing ECC efficiency is scalar multiplication~\cite{haddaji2024computing}, a fundamental operation in ECC-based protocols ~\cite{kumar2020design}. Efficient scalar multiplication reduces latency and power consumption, thereby facilitating faster and more secure data transmission.

One of the main challenges in Elliptic Curve Cryptography (ECC) lies in the high computational cost of scalar multiplication. As the scale and computational demands of modern information systems continue to grow~\cite{ullah2023elliptic}, the need for more efficient and secure scalar multiplication algorithms becomes increasingly critical~\cite{xie2023robust}. Existing approaches often struggle to balance computational speed and memory usage, particularly in resource-constrained environments.

To address these challenges, this paper proposes an accelerated scalar multiplication algorithm based on M-ary precomputation. The proposed method reduces the time complexity from $\Theta(Q \log p)$ to $\Theta\left(\frac{Q \log p}{\log Q}\right)$ while achieving flexible memory efficiency through structured precomputation. This work aims to enhance the performance and scalability of ECC operations across a broad range of cryptographic applications.

The key contributions of this research are as follows:

\begin{enumerate}
    \item \textbf{Innovative Scalar Multiplication Algorithm}:  
    We propose a flexible M-ary precomputation-based scalar multiplication algorithm for Elliptic Curve Cryptography (ECC), achieving significant improvements in both time and memory efficiency. Compared to conventional methods such as the sliding window and fixed-base comb algorithms, the proposed method achieves linear memory scaling and better adapts to arbitrary scalar sizes.

    \item \textbf{Theoretical Improvements in Time and Memory Complexity}:  
    Our method reduces the time complexity from $\Theta(Q \log p)$ to $\Theta\left(\frac{Q \log p}{\log Q}\right)$ and the memory complexity to $\Theta\left(\frac{Q \log p}{\log^2 Q}\right)$ through structured precomputation and sparse storage. This provides an asymptotic advantage over traditional precomputation techniques.

    \item \textbf{Significant Performance Gains}:  
    Experimental evaluations based on ElGamal encryption and NS3 communication simulations demonstrate the practical effectiveness of the proposed method. Specifically, the algorithm achieves up to a 59\% reduction in encryption time and a 30\% reduction in peak memory usage compared to baseline methods when $Q = 1000$. Moreover, the binary-optimized variant achieves a 22.1\% reduction in total communication time on secp384r1 and a 25.4\% reduction in overall simulation time on secp521r1, highlighting its scalability across different elliptic curves and workloads.
\end{enumerate}

\section{Related work and preliminaries}

\subsection{Related Work}
ECC has been extensively studied for its efficiency and security advantages, particularly in resource-constrained settings. Existing research can be broadly categorized into three areas: optimizations of ECC algorithms, applications of ECC in resource-constrained settings, and lightweight cryptographic techniques.

Scalar multiplication is the most computationally intensive operation in elliptic curve cryptography (ECC) and has been the subject of extensive optimization research~\cite{haddaji2024computing, kumar2020design}. Traditional methods such as Double-and-Add~\cite{bos2014elliptic} and NAF-based algorithms~\cite{avanzi2011redundant} focus on reducing the number of point additions. More advanced techniques, including $2^k$-ary methods~\cite{zhang2023data}, Sliding Window algorithms~\cite{rivain2011fast, yang2024sakms}, Montgomery Ladder~\cite{yang2024sakms, ansari2008high}, Fixed-Base Comb~\cite{mohamed2012improved}, and Window $\tau$-NAF representations~\cite{yu2021pre}, further enhance computational efficiency or side-channel resistance. While these approaches significantly improve scalar multiplication performance, they often encounter scalability and memory trade-offs when handling large numbers of scalar operations. This paper addresses these limitations by proposing an M-ary precomputation-based algorithm that simultaneously improves time complexity and memory efficiency, providing a flexible and scalable solution for ECC applications.

\begin{table}[h!]
\centering
\resizebox{\textwidth}{!}{ 
\begin{tabular}{|p{5.5cm}|p{3.5cm}|p{6cm}|}
\hline
\textbf{Research Area} & \textbf{Related Work} & \textbf{Key Contributions} \\ \hline
\multirow{4}{*}{ECC Algorithm Optimization} &
  Scalar Multiplication & Reduce the number of point additions and multiplications~\cite{bos2014elliptic, avanzi2011redundant} \\ \cline{2-3} 
 & Precomputation Techniques & Improve performance but require significant memory~\cite{rivain2011fast, yang2024sakms} \\ \cline{2-3} 
 & Homomorphic Encryption & Focus on high-performance computing\cite{cheon2023homomorphic} \\ \cline{2-3}
 & GPU-Accelerated Computations & Target high-performance computing rather than IoT~\cite{narisada2023gpu} \\ \hline
\multirow{4}{*}{\parbox{4.5cm}{ECC Applications \\ in Resource-constrained Settings}} & Secure Communication &
  EEC’s Error Correction and Side-Channel Attack Resistance~\cite{brier2002weierstrass} \\ \cline{2-3} 
 & Decentralized Federated Learning & Enhances security and scalability in IoT~\cite{yan2024performance} \\ \cline{2-3} 
 & Physical Layer Security & Improves data security in IoT~\cite{illi2023physical} \\ \cline{2-3} 
 & Machine Learning in ECC & Anomaly detection and authentication enhance ECC's robustness~\cite{fu2023modeling, al2020survey} \\ \hline
\multirow{4}{*}{Lightweight Cryptography} &
  Lightweight Protocols & Optimize resource utilization while maintaining strong encryption~\cite{ullah2023elliptic} \\ \cline{2-3} 
 & Tensor-Based Models & Enhance reliability and robustness in hybrid IoT systems~\cite{fan2023tensor} \\ \cline{2-3} 
 & Standard ECC Limitations & Standard ECC implementations struggle with IoT constraints~\cite{cao2024practical, patel2023ebake} \\ \cline{2-3} 
 & Lightweight ECC Protocols & Improve scalability and efficiency for IoT applications~\cite{cao2024input} \\ \hline
\end{tabular}
}
\caption{Summary of Related Work in ECC and IoT Applications}
\label{tab:related_work}
\end{table}

Recent advancements in ECC optimizations, including homomorphic encryption~\cite{cheon2023homomorphic} and GPU-accelerated computations~\cite{narisada2023gpu}, primarily focus on high-performance computing. However, these methods demand additional resources, making them less suitable for resource-constrained settings. Hardware-level optimizations\cite{zeghid2023speed} lack portability, whereas algorithm-level optimizations are more beneficial for practical applications.

ECC is widely employed in communications in resource-constrained settings because of its small key size and high computational efficiency~\cite{ye2024ecc}. High-impact applications include decentralized federated learning (DFL)~\cite{yan2024performance}, physical layer security~\cite{illi2023physical}, and secure multimedia transmission techniques, such as 3D scrambling~\cite{sahasrabuddhe2021multiple}.Machine learning methods, such as anomaly detection and authentication~\cite{fu2023modeling, al2020survey}, have further enhanced the robustness of ECC-based systems in resource-constrained settings.

Despite these advances, the computational limitations of resource-constrained settings frequently result in performance bottlenecks~\cite{xie2023robust, gao2023deep, barbarossa2023semantic}. Traditional ECC implementations face challenges in balancing security and real-time communication requirements in high-frequency IoT data exchanges~\cite{huo2024cluster, patel2023ebake}.

Lightweight cryptography has emerged as a promising solution to the efficiency challenges in resource-constrained settings~\cite{ullah2023elliptic}. These protocols optimize resource utilization while maintaining strong encryption, thereby ensuring scalability and robustness~\cite{cao2024input}. Techniques such as tensor-based models have been proposed to enhance network reliability and robustness in hybrid resource-constrained systems~\cite{fan2023tensor}.

However, existing lightweight cryptographic protocols typically rely on standard ECC implementations, which are not optimized for the specific constraints of resource-constrained devices~\cite{cao2024practical, patel2023ebake}. The need for fundamental improvements in ECC algorithms remains largely unaddressed, resulting in a significant gap in the literature.

Although ECC has undergone multiple optimizations, the computational cost remains unacceptable when a large number of scalar multiplications are required\cite{zhang2023high}.No existing method can identify the path with the shortest step length for a specific computational task while balancing computational and space costs. Hardware-optimized scalar multiplication demonstrates scalability and robustness in resource-constrained settings; however, its reliance on traditional methods limits efficiency~\cite{xie2023robust, gao2023deep, awaludin2021high}. Memory-intensive approaches, such as Sliding Window techniques~\cite{rivain2011fast, yang2024sakms}, are unsuitable for IoT devices with limited storage capacity. This study addresses these gaps by introducing a novel M-ary Precomputation-Based Accelerated Scalar Multiplication algorithm. By reducing both time complexity and memory usage, the proposed method offers a scalable and efficient solution for secure information communication.

\subsection{Preliminaries}
In this section, we present the core principles and operations related to elliptic curves, as well as the scalar multiplication algorithm and its optimization techniques.

\subsubsection{Elliptic Curves~\cite{aranha2023survey}}
An elliptic curve over a finite field $ \mathbb{F}_{p} $ is of this form:
\begin{multline}
E_p(a,b)=\begin{Bmatrix}(x,y) | x,y\in\mathbb{F}_p,\\
y^2\equiv x^3+ax+b(\operatorname{mod}p)\end{Bmatrix}\cup\begin{Bmatrix}O\end{Bmatrix}.
\end{multline}

Where $ p $ is a prime number, and $ a, b \in \mathbb{F}_p $ satisfy the condition $ 4a^3 + 27b^2 \not\equiv 0 \pmod{p} $, the point $ O $ is referred to as the point at infinity. When there is no need to specify $ a $ and $ b $, the elliptic curve $ E_{p}(a,b) $ can also be denoted as $ E(\mathbb{F}_p) $. The field $ \mathbb{F}_p $ is called the base field of the elliptic curve $ E(\mathbb{F}_p) $.

The process of adding points on an elliptic curve is defined as follows.

Let $ P = (x_1, y_1) $ and $ Q = (x_2, y_2) $ be points in $ E_p(a, b) \setminus \{O\} $. Then, the addition of $ P $ and $ Q $ is defined as follows:
\begin{align}
P+Q=\begin{cases}\quad O,&\quad x_1=x_2,y_1=-y_2,\\(x_3,y_3),&\quad\text{otherwise}.\end{cases}
\end{align}
where
\begin{align}
x_{3}=\lambda^{2}-x_{1}-x_{2},\\
y_{3}=\lambda\left(x_{1}-x_{3}\right)-y_{1},
\end{align}
\begin{align}
    \lambda=\begin{cases}\frac{y_2-y_1}{x_2-x_1},&P\neq Q,\\[2ex]\frac{3x_1^2+a}{2y_1},&P=Q.\end{cases}
\end{align}

\subsubsection{Scalar Multiplication Algorithms~\cite{ansari2008high}}
Let $ n \in \mathbb{N} $ and $ P \in E_p(a, b) $. Then, the scalar multiplication of the non-negative integer $ n $ with the point $ P $ is defined as follows:
\begin{equation}nP=\left\{\begin{matrix}O,&n=0,\\\left(n-1\right)P+P,&n\geq1.\end{matrix}\right.\end{equation}

Next, we introduce seven widely adopted optimization algorithms for elliptic curve scalar multiplication: the double-and-add algorithm~\cite{bos2014elliptic}, the NAF-based scalar multiplication algorithm~\cite{avanzi2011redundant}, the $2^{k}$-ary method~\cite{zhang2023data}, the Sliding Window algorithm~\cite{rivain2011fast, yang2024sakms}, the Montgomery Ladder technique~\cite{yang2024sakms, ansari2008high}, the Fixed-Base Comb algorithm~\cite{mohamed2012improved}, and the Window $\tau$-NAF precomputation scheme~\cite{yu2021pre}. These algorithms have been extensively employed to enhance the computational efficiency of scalar multiplication in elliptic curve cryptography (ECC).

Among them, the Sliding Window algorithm~\cite{rivain2011fast, yang2024sakms} significantly reduces the number of point additions by leveraging a precomputed table of odd multiples of the base point. The Montgomery Ladder algorithm~\cite{yang2024sakms, ansari2008high} offers a constant-time implementation that is inherently resistant to side-channel attacks, thus achieving a favorable trade-off between security and efficiency. The Fixed-Base Comb algorithm~\cite{mohamed2012improved} partitions the scalar into multiple columns based on a fixed base, enabling parallel computation and fast table lookups when the base point is constant. The Window $\tau$-NAF precomputation scheme~\cite{yu2021pre} combines the efficiency of non-adjacent form (NAF) representations with $\tau$-adic expansions in Koblitz curves, further reducing the number of required operations through strategic windowing and precomputation.

These algorithms are widely utilized to improve the efficiency and security of scalar multiplication in ECC. Table~\ref{tbl:Related Methods} summarizes the time and space complexities of these algorithms.

\begin{table}[!t]
    \caption{Comparative Analysis of Time and Space Complexities for Related Elliptic Curve Scalar Multiplication Algorithms.\label{tbl:Related Methods}}
    \centering
    \begin{tabular}{ccc}
    \toprule[1.5pt]
    Algorithm & Time Complexity & Space Complexity\\
    \midrule[1pt]
    Double-and-add ~\cite{bos2014elliptic} & $\Theta\left(Q\mathrm{log~}p\right)$ & $\Theta(1)$\\
    NAF-based ~\cite{avanzi2011redundant} & $\Theta\left(Q\mathrm{log~}p\right)$ & $\Theta(1)$\\
    $2^{k}$-ary ~\cite{zhang2023data}& $\Theta\left(Q\mathrm{log~}p\right)$ & $\Theta(1)$\\
    Sliding Window ~\cite{rivain2011fast, yang2024sakms}& $\Theta\left(\frac{Q\mathrm{log~}p}{r}\right)$ & $\Theta\left(2^r\right)$\\
    Montgomery Ladder ~\cite{rivain2011fast, yang2024sakms} & $\Theta\left(Q\mathrm{log~}p\right)$ & $\Theta(1)$\\
    Fixed-Base Comb~\cite{mohamed2012improved} & $\Theta\left(\frac{Q\mathrm{log~}p}{r}\right)$ & $\Theta\left(2^r\right)$\\
    Window $\tau$-NAF~\cite{yu2021pre} & $\Theta\left(\frac{Q\mathrm{log~}p}{r}\right)$ & $\Theta\left(2^r\right)$\\
    \bottomrule[1.5pt]
\end{tabular}
\end{table}

\subsubsection{Methods}

\noindent\textbf{Double-and-add Algorithm~\cite{bos2014elliptic}}. The Double-and-Add Algorithm is a fundamental method for scalar multiplication in elliptic curve cryptography (ECC). Its simplicity has significantly contributed to the development of scalar multiplication algorithms, with many subsequent algorithms being optimizations of this approach.

The core of Double-and-add Algorithm in elliptic curve cryptography is the following identity:
\begin{equation}
k=2\left\lfloor\frac{k}{2}\right\rfloor+k\operatorname{mod}2.
\end{equation}

This transformation reduces the computation of $kP$ to that of $\left\lfloor\frac{k}{2}\right\rfloor P$. Consequently, the time complexity $T(k)$ for computing the scalar multiplication $kP$ follows the recursive formula:
\begin{equation}
T(k)=T\left(\frac{k}{2}\right)+\Theta(1).
\end{equation}

\noindent\textbf{NAF Based Scalar Multiplication Algorithm~\cite{avanzi2011redundant}}. The Double-and-Add Algorithm~\cite{bos2014elliptic} is a basic yet effective method for scalar multiplication; however, its performance can often be improved. This has led to the development of scalar multiplication algorithms based on the Non-Adjacent Form (NAF)~\cite{avanzi2011redundant}, which optimize the computation through a novel representation of integers.

The NAF representation expresses the scalar $k$ in a way that minimizes the number of required point additions. Specifically, to compute the scalar multiplication $kP$, we represent $k$ in its NAF as:
\begin{equation}
    k = \sum_{i=0}^{d-1} a_i 2^i.
\end{equation}

The scalar multiplication is then computed as:
\begin{equation}
    kP = \sum_{i=0}^{d-1} a_i (2^i P).
\end{equation}

Denoting  $A_{i}=2^{i}P$ for $i \geq 0$, the scalar multiplication can be rewritten as:
\begin{equation}
    kP = \sum_{i=0}^{d-1} a_i A_i.
\end{equation}

The NAF representation typically results in fewer elliptic curve point additions compared to the Double-and-Add method, leading to improved practical efficiency in scalar multiplication.\newline

\noindent\textbf{\boldmath{$2^{k}$}-ary~\cite{zhang2023data}}. The $2^k$-ary algorithm~\cite{zhang2023data} is used to compute scalar multiplication $mP$. The algorithm begins by selecting a positive integer $k$ and representing the scalar coefficient $m$ as a base-$2^k$ number:

Where $0 \leq a_i < 2^k$ and $a_{d-1}>0$,the algorithm defines the intermediate values $A_j$ as follows:
\begin{equation}
A_j:=\sum_{i=d-j}^{d-1}a_i2^{k\left(i-(d-j)\right)}P,0\leq j\leq d.
\end{equation}

With $A_0 = O$ and $A_d = mP$. For $1 \leq j \leq d$, we have:
\begin{equation}
A_j=2^kA_{j-1}+a_{d-j}P.
\end{equation}

To simplify the computation, each coefficient $a_i$ can be expressed as:
\begin{equation}
a_{i}=u_i\cdot 2^{s_i},
\end{equation}
where $u_i$ is an odd integer (i.e., $u_i$ is not divisible by 2), and $s_i$ is a non-negative integer that represents the power of 2 in the factorization of $a_i$.

The steps involve doubling $A_{j-1}$ a total of $k - s_{d-j}$ times to compute $2^{k - s_{d-j}} A_{j-1}$, retrieving $u_{d-j}P$ from a precomputed table, and adding this to the doubled result. The final result is then doubled $s_{d-j}$ times to obtain $A_j$.

The advantage of this method lies in reducing the number of point additions (excluding doublings) required on the elliptic curve, thereby enhancing efficiency. In practice, setting $k = 5$ is often optimal for performance.\newline

\noindent\textbf{Sliding Window algorithm~\cite{rivain2011fast, yang2024sakms}}. The Sliding Window algorithm~\cite{rivain2011fast, yang2024sakms} is an optimization technique in elliptic curve cryptography (ECC) designed to enhance the efficiency of scalar multiplication, computing $Q = kP$, where $k$ is the scalar and $P$ is a point on the elliptic curve. This algorithm partitions $k$ into fixed-size windows of $\omega$ bits, with each window $w_i$ representing a segment of the scalar.

The algorithm processes $k$ in $\omega$-bit windows by precomputing and storing each odd multiple $(2j + 1)P$ in a table, which reduces redundant computations. During execution, for a given window $w_i$, the algorithm combines point doubling and table lookups, performing the operation:
\begin{equation}
    Q \gets 2^{\omega} Q + P_{w_i},
\end{equation}
where $P_{w_i}$ is the precomputed value for the window $w_i$, and $2^{\omega} Q$ represents $\omega$ consecutive doublings. This approach significantly reduces the number of point additions, balancing memory usage with computational complexity.

The Sliding Window algorithm~\cite{rivain2011fast, yang2024sakms} requires $O(2^{\omega-1})$ storage for precomputed values and reduces the average number of point additions from $\lfloor \log_2 k \rfloor$ (as seen in the Double-and-Add algorithm) to approximately $\lfloor \log_2 k \rfloor / \omega$. By tuning $\omega$, the algorithm can adapt to various performance and hardware constraints, making it particularly effective for scalars with large bit lengths.\newline

\noindent\textbf{Montgomery Ladder algorithm ~\cite{ansari2008high}}. The Montgomery Ladder algorithm~\cite{ansari2008high} is an efficient and secure method for scalar multiplication in elliptic curve cryptography (ECC), particularly noted for its resistance to side-channel attacks. This algorithm computes the scalar multiplication $Q = kP$ by maintaining a consistent operation flow, regardless of the scalar $k$, which mitigates simple power analysis (SPA) vulnerabilities.

In the Montgomery Ladder, the scalar $k$ is processed bit-by-bit from the most significant to the least significant bit. At each iteration, two points, $R_0$ and $R_1$, are updated to satisfy the invariant $ R_1 - R_0 =P$. This ensures that only differential point additions and doublings are used, enhancing security. 

The algorithm is compatible with various elliptic curve coordinate systems, including Jacobian and projective coordinates, optimizing performance based on specific hardware or software constraints. Its robustness against side-channel attacks, along with its adaptability to different elliptic curve forms, underscores its importance in modern ECC implementations.

The operational flow of Sliding Window algorithm and Montgomery Ladder algorithm is illustrated in Algorithm~\ref{alg:sliding_window} and Algorithm~\ref{alg:montgomery_ladder}, respectively.





\makebox[\textwidth]{%
  \begin{minipage}[t]{0.60\textwidth}
  \centering
  \scriptsize
  \noindent\begin{algorithm}[H]
  \caption{Sliding Window Algorithm}
  \label{alg:sliding_window}
  \begin{algorithmic}
  \STATE \textbf{Input: } $P \in E(\mathbb{F}_q)$, $k = \sum_{i=0}^{l-1} k_i 2^i$, window size $\omega$
  \STATE \textbf{Output: } $Q = kP$
  \STATE Precompute $P_{2i+1}$, $Q \gets O$, $i \gets l-1$
  \WHILE{$i \geq 0$}
    \IF{$k_i = 0$}
      \STATE $Q \gets 2Q$, $i \gets i-1$
    \ELSE
      \STATE Find max $t$ s.t. $i - t + 1 \leq \omega$, $k_t = 1$
      \STATE $h_i \gets (k_i \dots k_t)_2$
      \STATE $Q \gets [2^{i-t+1}]Q + P_{h_i}$
      \STATE $i \gets t - 1$
    \ENDIF
  \ENDWHILE
  \STATE \textbf{Return: } $Q$
  \end{algorithmic}
  \end{algorithm}
  \end{minipage}
  \hfill
  \begin{minipage}[t]{0.40\textwidth}
  \centering
  \scriptsize
  \noindent\begin{algorithm}[H]
  \caption{Montgomery Ladder Algorithm}
  \label{alg:montgomery_ladder}
  \begin{algorithmic}
  \STATE \textbf{Input: } $P \in E(\mathbb{F}_q)$, scalar $k = \sum_{i=0}^{n-1} k_i 2^i$
  \STATE \textbf{Output: } $Q = kP$
  \STATE $R_0 \gets O$, $R_1 \gets P$
  \FOR{$i = n-1$ downto $0$}
    \IF{$k_i = 0$}
      \STATE $R_1 \gets R_0 + R_1$
      \STATE $R_0 \gets 2R_0$
    \ELSE
      \STATE $R_0 \gets R_0 + R_1$
      \STATE $R_1 \gets 2R_1$
    \ENDIF
  \ENDFOR
  \STATE \textbf{Return: } $R_0$
  \end{algorithmic}
  \end{algorithm}
  \end{minipage}%
}
\newline

\noindent\textbf{Fixed-Base Comb algorithm~\cite{mohamed2012improved}}. The Fixed-Base Comb algorithm~\cite{mohamed2012improved} represents the scalar $k$ in width-$\omega$ Non-Adjacent Form (NAF) and divides it into $\omega \times v$ blocks, processed from top to bottom and right to left.

First, the scalar $k$ is divided into $a = \lceil l / \omega \rceil $ blocks of size $\omega$, with padding added if necessary. Each block $K_d$ consists of $\omega$ bits, allowing $k$ to be expressed as a series of these blocks. The scalar multiplication $kP$ can thus be computed as follows:
\begin{equation}
kP = K_{a-1}K_{a-2} \ldots K_0 P = \sum_{j=0}^{v-1} \sum_{t=0}^{b-1} (K_{jb+t} 2^{t \omega}) P,
\end{equation}
where $K_{jb+t}$ represents the block in width-$\omega$ NAF representation.

To optimize performance, values $G[j][sd]$ are precomputed for necessary indices, allowing $kP$ to be expressed in a more efficient form:
\begin{equation}
kP=\sum_{t=0}^{b-1}2^{t\omega}(\sum_{j=0}^{v-1}G[j][I_j,t]).   
\end{equation}

This algorithm effectively balances the average and worst-case densities of non-zero digits in the width-$\omega$ NAF representation, contributing to its overall efficiency. By optimizing the representation and processing of the scalar $k$, this approach significantly enhances the performance of scalar multiplication in elliptic curve cryptography.

\noindent\textbf{Pre-computation Scheme of Window $\tau$ NAF~\cite{yu2021pre}}. Yu and Xu revisit the pre-computation scheme for the window $\tau$ NAF (non-adjacent form) method, specifically designed for Koblitz curves. Their innovative approach enhances the efficiency of scalar multiplication by introducing new algebraic operations known as $\mu \bar{\tau }$-operations~\cite{yu2021pre}, which utilize the complex conjugate of the Frobenius map. This improvement surpasses traditional methods, resulting in fewer field operations during the pre-computation phase and facilitating faster scalar multiplication.

This algorithm uses a unified pre-computation scheme that maintains efficiency across different configurations of Koblitz curves, particularly for curves $E_0$ and $E_1$. Their results reveal time complexities of $6M + 6S$, $18M + 17S$, and $44M + 32S$ for window widths of 4, 5, and 6, respectively, when $a = 0$. These costs are approximately twice as efficient compared to existing pre-computation techniques.

Overall, this innovative pre-computation method not only enhances computational efficiency but also suggests the potential for further advancements in elliptic curve cryptography, particularly for other types of curves defined over $F_3^m$ or $F_{q^m}$ for primes $q \geq 5$. Thus, their work significantly contributes to both theoretical understanding and practical implementations in the field.

\section{Our Proposed Algorithm}

In this section, we present an optimized approach for scalar multiplication based on M-ary precomputation. This method is designed to accelerate the computation of multiple scalar multiplications on elliptic curves, significantly improving performance over previous algorithms. We will outline the key principles behind the algorithm, its time complexity, and the optimizations that minimize computational overhead. Recent advances in the theoretical understanding of time-space tradeoffs for function inversion~\cite{golovnev2023revisiting} suggest that careful balancing of precomputation storage and online computation cost is crucial for achieving optimal performance. Motivated by these insights, our M-ary precomputation-based method is designed to optimize both memory consumption and computational efficiency in structured scalar multiplication tasks.

\begin{figure}[H]
    \centering
    \captionsetup{justification=centering,singlelinecheck=false}
    \includegraphics[width=0.80\textwidth]{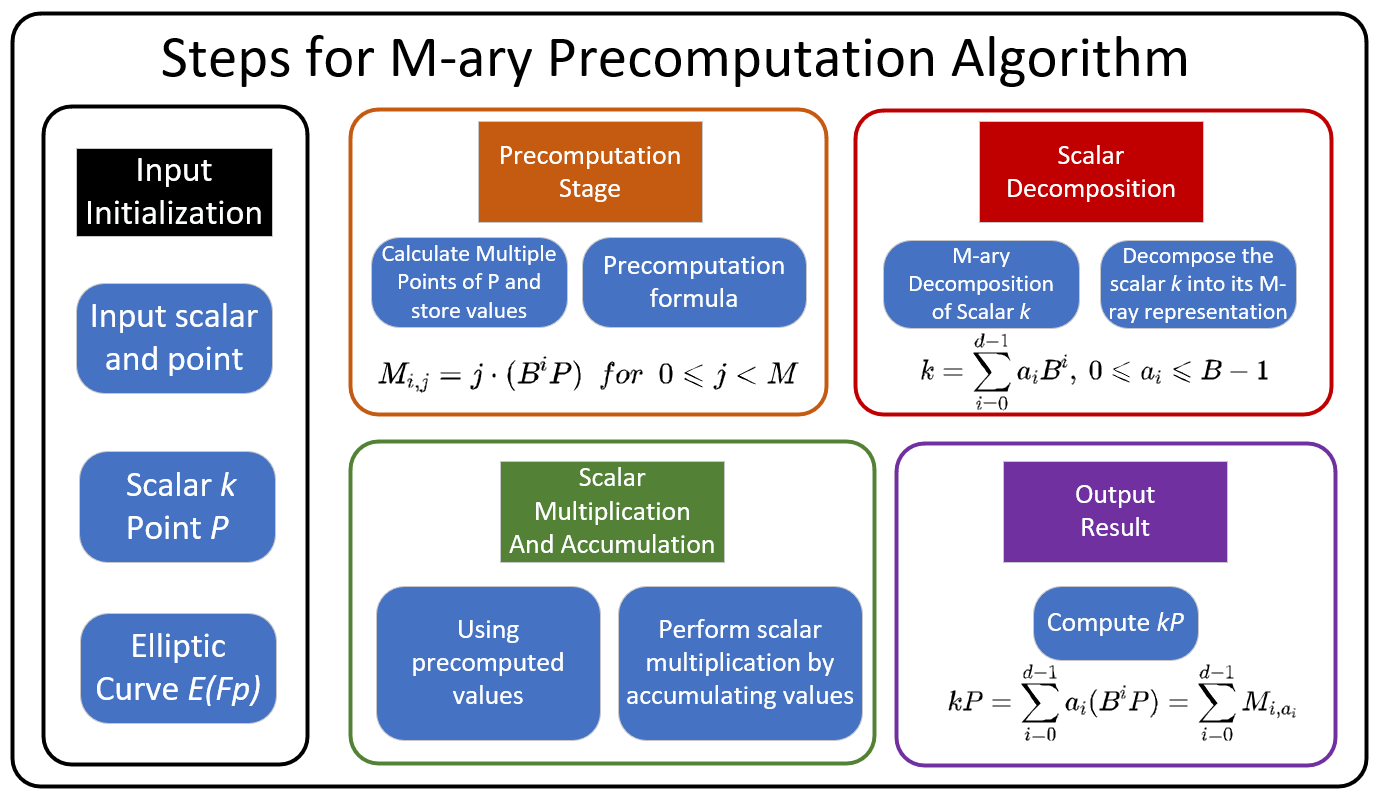}
    \caption{Steps for M-ary Precomputation-Based Algorithm, illustrating the process from input initialization, precomputation, scalar decomposition, scalar multiplication, and the final result computation.}
    \label{fig:mary_steps}
\end{figure}

\begin{algorithm}[H]
\footnotesize
\caption{M-ary Precomputation-Based Scalar Multiplication}\label{alg:mary}
\begin{algorithmic}
\STATE \textbf{Input: } Scalars $k_1, \dots, k_Q \in \mathbb{N}$; base point $P \in E_p(a,b)$
\STATE \textbf{Output: } $\left(k_1P, k_2P, \dots, k_QP\right)$
\vspace{0.3em}

\STATE // Step 1: Determine parameters $d$ and $B$
\STATE $d \gets \left\lceil \frac{\ln p}{W(Q/e) + 1} \right\rceil$
\STATE $B \gets \left\lceil \sqrt[d]{n} \right\rceil$

\STATE // Step 2: Precompute $M[i][j]$ for $i \in [0,d-1]$, $j \in [0,B]$
\FOR{$i = 0$ to $d - 1$}
  \FOR{$j = 0$ to $B$}
    \STATE Compute $M[i][j] \gets j \cdot B^i \cdot P$
    \STATE Verify $M[i][j]$ satisfies the elliptic curve equation // \textit{FI protection}
  \ENDFOR
\ENDFOR

\STATE // Step 3: For each scalar $k$, compute $kP$
\FOR{each $k \in \{k_1, \dots, k_Q\}$}
  \STATE Choose random $r$ and shift $k \gets k + r \cdot B^d$
  \STATE Decompose $k$ into base-$B$: $k = \sum_{i=0}^{d-1} k_i B^i$
  \STATE Compute $S \gets \sum_{i=0}^{d-1} M[i][k_i]$
  \STATE Correct: $S \gets S - r \cdot (B^d \cdot P)$
  \STATE Store $S$
\ENDFOR

\STATE \textbf{return} all computed $k_iP$
\end{algorithmic}
\end{algorithm}

\Cref{fig:mary_steps} provides a visual breakdown of each phase in the M-ary precomputation based algorithm. Additionally, \Cref{alg:mary} offers a comprehensive explanation of the computational workflow for the M-ary precomputation-based accelerated scalar multiplication algorithms.

\subsection{Overview of the Problem}

Scalar multiplication on elliptic curves is a fundamental operation in many cryptographic protocols. Given a point $ P $ on an elliptic curve $ E_p(a, b) $, the goal is to compute $ kP $ for a scalar $ k $. Traditional methods for scalar multiplication suffer from time complexities proportional to $ \Theta(Q \log p) $, where $ Q $ is the number of multiplications and $ p $ is the size of the finite field. This motivates the search for algorithms that can achieve lower time complexities.

\subsection{The M-ary Precomputation Method}

Our method leverages the fixed nature of both the elliptic curve and the base point $ G $. Given that the same scalar multiplication is repeatedly computed, we utilize the M-ary precomputation technique to precompute and store multiples of the base point.

\subsubsection{Scalar Representation}

We represent the scalar $ k $ as a sum of powers of a base $ B $:
\begin{equation}
k = \sum_{i=0}^{d-1} a_i B^i, \quad 0 \leq a_i \leq B - 1.
\end{equation}

This allows the scalar multiplication to be expressed as:
\begin{equation}
kP = \sum_{i=0}^{d-1} a_i \left( B^i P \right),
\end{equation}
where each term $ B^i P $ is precomputed and stored in a table $ M $, which significantly reduces the computational overhead when performing scalar multiplications.

\subsubsection{Precomputing the Table $ M $}

The table $ M $ is defined as:
\begin{equation}
M_{i,j} = j \left( B^i P \right), \quad 0 \leq i \leq d-1, \quad 0 \leq j \leq B.
\end{equation}

This table can be precomputed with a time complexity of $ \Theta(dB) $. By using the recurrence relation:
\begin{equation}
M_{i,j} = M_{i,j-1} + M_{i,1}, \quad j \geq 2,
\end{equation}
and the base case:
\begin{equation}
M_{i,1} = B^i P,
\end{equation}
we can efficiently fill the table for all values of $ i $ and $ j $. Once precomputation is done, the scalar multiplication $ kP $ can be quickly computed by summing the appropriate values from the table $ M $:
\begin{equation}
kP = \sum_{i=0}^{d-1} M_{i, a_i}.
\end{equation}

Thus, the time complexity for computing a single scalar multiplication is $ \Theta(d) $, and for computing $ Q $ scalar multiplications, it becomes $ \Theta(dQ) $.

\subsection{Optimizing Time Complexity}

To further optimize the algorithm, we seek to minimize the total computational cost by selecting the appropriate parameters for $ d $ and $ B $. The relationship between $ B $ and $ d $ is derived from the requirement that $ B^d - 1 \geq n - 1 $, where $ n $ is the maximum possible value for the scalar $ k $.

Since the time complexity of the algorithm is $ \Theta(d(B + Q)) $, we aim to minimize the expression by selecting $ B $ that balances both the base $ B $ and the number of precomputed values. Through mathematical analysis (see Appendix B for detailed proofs), the optimal value of $ B $ is found to be:
\begin{equation}
B^* = \left\lceil \sqrt[d]{n} \right\rceil.
\end{equation}

Substituting this into the time complexity expression gives us:
\begin{equation}
\Theta\left(d \left( \sqrt[d]{n} + Q \right)\right.
\end{equation}

\subsection{Minimizing the Parameter $ d $}

To determine the optimal value of $ d $, we seek the minimum of the function:
\begin{equation}
f(x) = x \left( p^{\frac{1}{x}} + Q \right), \quad x > 0.
\end{equation}

By differentiating $ f(x) $ and solving for the critical point, we find that the optimal value of $ d $ occurs at:
\begin{equation}
x_0 = \frac{\ln p}{W\left(\frac{Q}{\mathrm{e}}\right) + 1},
\end{equation}
where $ W(\cdot) $ is the principal branch of the Lambert W function. This value of $ d $ minimizes the overall time complexity, leading to an optimal trade-off between the precomputation cost and the number of scalar multiplications.
Thus, $ Q = O(f(x_0)) $. Since $ f(x_0) \leq f(d) < f(x_0) + \Theta(Q) $, we conclude:
\begin{equation}
f(d) = \Theta\left(f(x_0)\right) = \Theta\left(\frac{Q \ln p}{W\left( \frac{Q}{\mathrm{e}} \right)}\right) = \Theta\left(\frac{Q \log p}{\log Q}\right).
\end{equation}

\subsection{Space Complexity}

The spatial complexity arises from the precomputation of the table $ M $, where $ M_{i,j} = j(B^i P) $, for $ 0 \leq i \leq d-1 $ and $ 0 \leq j \leq B $.

When $ B = \left\lceil \sqrt[d]{n} \right\rceil $, $ x_0 = \frac{\ln p}{W\left( \frac{Q}{\mathrm{e}} \right)+1} $, and $ d = \left\lceil x_0 \right\rceil $, the spatial complexity is:
\begin{equation}
\Theta(dB) = \Theta\left(\frac{\ln p}{W\left(\frac{Q}{\mathrm{e}}\right)+1} \cdot p^{\frac{W\left(\frac{Q}{\mathrm{e}}\right)+1}{\ln p}}\right).
\end{equation}

Simplifying the exponential term, we get:
\begin{equation}
\Theta\left(\frac{\ln p}{W\left(\frac{Q}{\mathrm{e}}\right)+1} \cdot \exp\left(W\left(\frac{Q}{\mathrm{e}}\right)+1\right)\right).
\end{equation}

Since $ \exp\left(W\left(\frac{Q}{\mathrm{e}}\right)+1\right) = \mathrm{e} \cdot \frac{Q}{W\left(\frac{Q}{\mathrm{e}}\right)} $, the spatial complexity becomes:
\begin{equation}
\Theta\left(\frac{\ln p}{W\left(\frac{Q}{\mathrm{e}}\right)+1} \cdot \frac{Q}{W\left(\frac{Q}{\mathrm{e}}\right)}\right).
\end{equation}

Thus, the final expression simplifies to:
\begin{equation}
\Theta\left(\frac{Q \log p}{\log^2 Q}\right).
\end{equation}

\subsection{The M-ary Precomputation Method with Less Space}

When the number of input scalar multiplications $ Q $ is large, storing the precomputed points in table $ M $ with $ \Theta\left(\frac{Q \log p}{\log^2 Q}\right) $ memory overhead incurs significant memory costs, which may diminish the practical advantages of our method. The challenge of managing precomputed data efficiently in scalar multiplication has been emphasized in recent works like Elastic MSM~\cite{elastic_msm}, which advocate elastic and modular preprocessing to adapt to memory constraints. 

Building upon similar motivations, we design a binary sparse storage format for M-ary precomputation, significantly minimizing the number of stored points without compromising computational efficiency.
In this approach, the points stored in each row are represented by a few 'indices' that span between powers of two. Specifically, we construct the storage format as follows:
\begin{equation}
\left\{ M_{i,2^{0}}, M_{i,2^{1}}, M_{i,2^{2}},..., M_{i,2^{\log_{2}{B}}} \right\}, \quad 1 \leq i \leq d.
\end{equation}

This method allows us to perform binary decomposition for each $M_{i,j}$, thus reducing the required storage space.

\subsubsection{Bisection-Based Storage for Table M}

We express $ a_{i} $ as the sum of powers of 2:
\begin{equation}
a_{i} = \sum_{j=0}^{\left\lfloor \log_{2}{B} \right\rfloor} b_{j} 2^{j}, \quad kP = \sum_{i=0}^{d-1}\sum_{j=0}^{\left\lfloor \log_{2}{B} \right\rfloor} b_{j} 2^{j}  \left( B^i P \right).
\end{equation}

First, we compute the complete precomputation table $M$ using the conventional method, in which the total number of additions is approximately $O(d \cdot B)$, generating all elements of $M_{i,j}$. Subsequently, we select the bisection points from table $M$ and store them in a compact table $H$:
\begin{equation}
H_{i,j} = M_{i,2^{j}}, \quad 1 \leq i \leq d, \quad 1 \leq j \leq \left\lfloor \log_{2}{B} \right\rfloor.
\end{equation}

We propose storing only table $H$, which enables efficient computation of $kP$ by selectively retrieving and combining the appropriate precomputed points from $H$ during scalar multiplication:
\begin{equation}
kP = \sum_{i=0}^{d-1}\sum_{j=0}^{\left\lfloor \log_{2}{B} \right\rfloor} b_{j} H_{i, j}.
\end{equation}

\subsubsection{Determination of Parameter $ d $}

Since the complete table $ M $ must be computed during precomputation, the time complexity is $ \Theta(dB) $. After adopting bisection-based storage, each digit $ a_i $ requires $ \log_{2}{B} $ scalar additions, resulting in a time complexity of $ \Theta(d\log_{2}{B}) $ for $ Q $ scalar multiplications. The overall time complexity is as follows:
\begin{equation}
\Theta\left(d\left(\sqrt[d]{p} + Q\log{\sqrt[d]{p}}\right)\right).
\end{equation}

By minimizing this time complexity, we can derive the optimal parameter $ d $ for the bisection-based storage scheme as:
\begin{equation}
d=\log_{2}{p},\quad B=\sqrt[d]{p} = e.
\end{equation}

\subsubsection{Complexity Analysis}

We analyze the algorithm's complexity in terms of space and time requirements. Let $d$ denote the decomposition depth and $B$ the base used in the $B$-ary representation, with $p$ being the bit-length of the field and $Q$ the number of scalars.

\textbf{Space Complexity.} The algorithm maintains a precomputed table $M$ of size $d \times B$, resulting in a space complexity of:
\begin{equation}
\label{eq:space-complexity}
\Theta(dB) = \Theta(e \log p).
\end{equation}

\textbf{Time Complexity.} For each scalar multiplication, the algorithm performs $d$ additions, each involving a constant-time table lookup. For $Q$ scalar multiplications, the total time complexity becomes:
\begin{equation}
\label{eq:time-complexity}
\Theta(dQ \log B) = \Theta(Q \log p + e \log p) = \Theta(Q \log p).
\end{equation}

\textbf{Storage of Binary Elements.} Each row stores a bounded number of intermediate points, typically logarithmic in $B$, leading to an auxiliary storage complexity of $\Theta(\log p)$.

Overall, the algorithm achieves near-linear time scaling in $Q$ and logarithmic overhead in both precomputation and storage. Despite the reduced storage overhead, the binary storage scheme introduces additional computational cost during scalar multiplication, as reconstructing the required values from the compact table $H$ involves more additive operations. Therefore, this optimization is particularly suitable for deployment on resource-constrained devices where memory is limited, but additional computation is acceptable during runtime.

\subsection{Algorithm Security Analysis}

This section analyzes the security properties of the proposed M-ary precomputation based scalar multiplication algorithm from a theoretical perspective, focusing on its resistance to side-channel threats, fault injection vulnerabilities, and overall cryptographic soundness.

\subsubsection{Side-Channel Resilience}

To mitigate side-channel threats such as Simple Power Analysis (SPA) and Differential Power Analysis (DPA), the proposed algorithm incorporates several defensive mechanisms. First, all point additions are performed using a constant-time addition subroutine to prevent information leakage through timing variations. Second, a randomized scalar blinding technique is applied by transforming the scalar $k$ to $k + r \cdot B^d$, introducing entropy across executions. Finally, table lookups during scalar decomposition follow a regular access pattern without branching, thereby avoiding conditional data-dependent operations. 

These strategies collectively provide a baseline level of protection against timing-based and power-based leakage~\cite{kocher1996timing,brier2002weierstrass}. However, for deployment in highly adversarial environments, further enhancements such as unified point addition formulas or dummy table accesses may be required to strengthen resistance against advanced power analysis.

\subsubsection{Fault Injection Vulnerabilities}

Fault injection (FI) attacks pose a different class of threats~\cite{boneh1997importance}, where adversaries may attempt to manipulate internal states or induce faults in intermediate computations, particularly during table precomputation or scalar decomposition. To address this risk, the algorithm introduces a lightweight consistency check during the precomputation phase, where each point $M[i][j]$ is verified to satisfy the elliptic curve equation $y^2 = x^3 + ax + b$.

Despite this verification step, the algorithm currently lacks built-in redundancy or self-correction mechanisms. If a fault alters the value of a precomputed point or disrupts scalar decomposition, the algorithm may output incorrect results without detection. Future work could explore fault-resilient enhancements such as result consistency checks, parity-based encoding, or lightweight checksum techniques that preserve performance while improving robustness against active attacks.

\subsubsection{Theoretical Soundness}

From a theoretical standpoint, the algorithm ensures correctness in scalar multiplication, preserving group law properties on elliptic curves. Under the assumption that elliptic curve operations are implemented securely, the algorithm maintains indistinguishability under chosen plaintext attacks (IND-CPA), which is the foundational requirement for elliptic curve-based encryption schemes.

However, practical instantiations may still leak information through side channels or fault manipulation unless complemented by secure hardware support or cryptographic protocol-level protections. For applications in digital signatures, secure multiparty computation, or zero-knowledge settings, additional safeguards such as verifiable computation or range proofs may be necessary.

Table~\ref{tbl:security-comparison} summarizes the resistance of several representative scalar multiplication algorithms against common cryptographic attacks, including Simple Power Analysis (SPA), Differential Power Analysis (DPA), and Fault Injection (FI). As shown, our proposed M-ary algorithm achieves strong SPA resistance through constant-time computation and partial mitigation against DPA and FI via scalar randomization and lightweight consistency checks. Compared to classical methods such as double-and-add or sliding window, the proposed design offers improved baseline protection while maintaining lightweight efficiency.

\begin{table}[!t]
\caption{Comparative Security Features of Scalar Multiplication Algorithms\label{tbl:security-comparison}}
\centering
\footnotesize
\begin{tabular}{l@{\hspace{0.4em}}c@{\hspace{0.4em}}c@{\hspace{0.4em}}c}
\toprule[1.5pt]
\textbf{Algorithm} & \textbf{SPA Resistance} & \textbf{DPA Resistance} & \textbf{FI Resistance} \\
\midrule[1pt]
Double-and-Add & $\times$ & $\times$ & $\times$ \\
NAF-based & $\triangle$ & $\times$ & $\times$ \\
$2^k$-ary & $\triangle$ & $\times$ & $\times$ \\
Sliding Window & $\triangle$ & $\times$ & $\times$ \\
Montgomery Ladder & $\checkmark$ & $\triangle$ & $\times$ \\
Fixed-Base Comb & $\triangle$ & $\times$ & $\times$ \\
Pre-computation Scheme & $\triangle$ & $\times$ & $\times$ \\ 
\textbf{M-ary (ours)} & $\checkmark$ & $\triangle$ & $\triangle$ \\
\bottomrule[1.5pt]
\end{tabular}

\vspace{0.4em}
\footnotesize
\noindent
$\checkmark$: Strong built-in resistance \quad
$\triangle$: Partial mitigation \quad
$\times$: No inherent protection
\end{table}

\section{Evaluation}

In this section, we present an in-depth evaluation of the proposed M-ary precomputation-based accelerated scalar multiplication algorithm. The evaluation is conducted in three main phases:

First, we compare the theoretical algorithmic efficiency of our method against traditional scalar multiplication optimization algorithms, including Double-and-Add, NAF-based, and $ 2^k$-ary algorithms. This comparison highlights the potential efficiency gains of the proposed method from a computational complexity perspective.

Second, we evaluate the performance of various scalar multiplication optimization algorithms within the ElGamal cryptosystem. By incorporating each algorithm into the ElGamal encryption scheme and performing scalar multiplications over elliptic curves secp256k1, secp384r1, and secp521r1, we measure improvements in encryption and decryption times. The results demonstrate that the M-ary precomputation-based algorithm significantly accelerates cryptographic operations compared to other methods. This analysis underscores the practical applicability of our proposed algorithm in real-world cryptographic systems, showcasing its effectiveness beyond theoretical models.

Lastly, we conduct a series of NS3-based simulations to evaluate the practical performance impact of the proposed algorithm within a representative communication network framework. The simulated environment consists of five interconnected nodes, and the evaluation focuses on key metrics such as total encryption time, entire communication time, and overall simulation time. The results demonstrate that the M-ary precomputation-based algorithm effectively reduces encryption time and improves communication efficiency within the simulated network environment.

These evaluations encompass theoretical efficiency analysis, practical simulation experiments, and cryptographic performance testing. Together, they provide a comprehensive validation of the advantages offered by the M-ary precomputation-based algorithm in the context of elliptic curve cryptography.

\subsection{Evaluation on Theoretical Algorithmic Efficiency}

In the previous section, we analyzed the complexities of various scalar multiplication algorithms, including the Double-and-Add, NAF-based, $2^k$-ary, Sliding Window, Montgomery Ladder, Fixed-Base Comb, Fixed-Window, and M-ary precomputation-based algorithms. As shown in \Cref{tbl:Complexities}, the M-ary algorithm achieves superior time complexity compared to conventional methods, which generally offer advantages in space complexity. In many practical cryptographic applications, space complexity remains within acceptable bounds, making time efficiency the primary criterion for evaluating the performance of scalar multiplication algorithms. Therefore, our M-ary precomputation-based approach is theoretically more effective for real-world deployment.

\begin{table}[!t]
    \caption{Comparative Analysis of Time and Space Complexities for Elliptic Curve Scalar Multiplication Algorithms.\label{tbl:Complexities}}
    \centering
    \footnotesize
    \begin{tabular}{ccc}
    \toprule[1.5pt]
    Algorithm & Time Complexity & Space Complexity\\
    \midrule[1pt]
    Double-and-add & $\Theta\left(Q\mathrm{log~}p\right)$ & $\Theta(1)$\\
    NAF-based & $\Theta\left(Q\mathrm{log~}p\right)$ & $\Theta(1)$\\
    $2^{k}$-ary & $\Theta\left(Q\mathrm{log~}p\right)$ & $\Theta(1)$\\
    Sliding Window & $\Theta\left(\frac{Q\mathrm{log~}p}{r}\right)$ & $\Theta\left(2^r\right)$\\
    Montgomery Ladder & $\Theta\left(Q\mathrm{log~}p\right)$ & $\Theta(1)$\\
    Fixed-Base Comb & $\Theta\left(\frac{Q\mathrm{log~}p}{r}\right)$ & $\Theta\left(2^r\right)$\\
    Window $\tau$-NAF & $\Theta\left(\frac{Q\mathrm{log~}p}{r}\right)$ & $\Theta\left(2^r\right)$\\
    \textbf{M-ary (ours)} & $\Theta\left(\frac{Q\log p}{\log Q}\right)$ & $\Theta\left(\frac{Q\log p}{\log^2 Q}\right)$\\
    M-ary (binary) & $\Theta(Q \log p)$ & $\Theta(log p)$ \\
    \toprule[1.5pt]
\end{tabular}
\end{table}

\subsection{Evaluation on Quantitative Algorithmic Efficiency}

\subsubsection{Evaluation Setup}
To investigate the performance comparison of the four scalar multiplication algorithms mentioned above, we have selected three elliptic curves (secp256k1, secp384r1, secp521r1) defined by the Standards for Efficient Cryptography Group for testing purposes. The parameter quintuples $ T = (p, a, b, G, n) $ for each elliptic curve are provided in the Appendix.



\subsubsection{Evaluation on the Efficiency of Scalar Multiplication Algorithms}
\begin{figure*}[!t]
\centering
\subfloat[]{\includegraphics[width=0.3\textwidth]{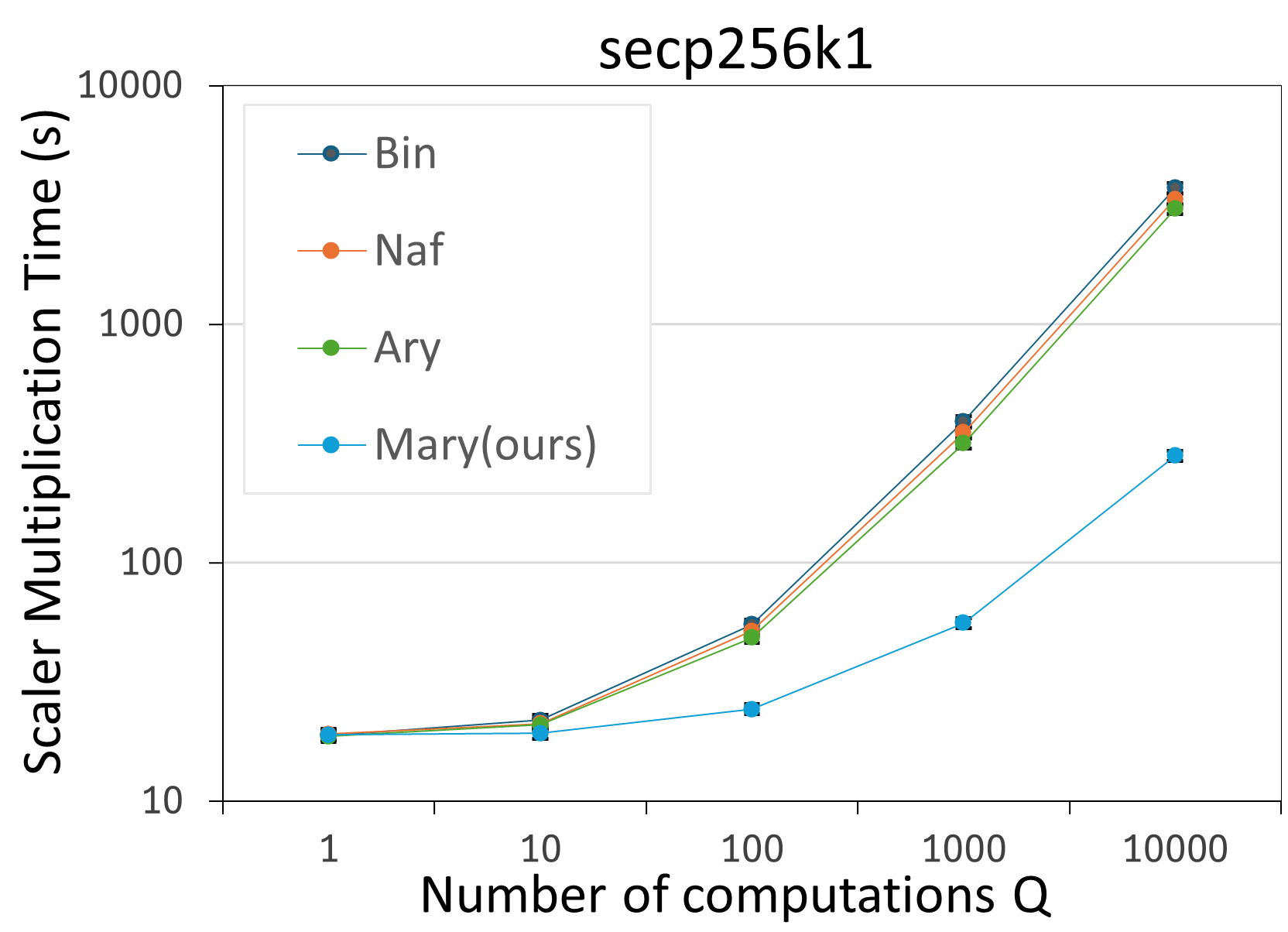}%
\label{fig:exp1a}}
\hfil
\subfloat[]{\includegraphics[width=0.3\textwidth]{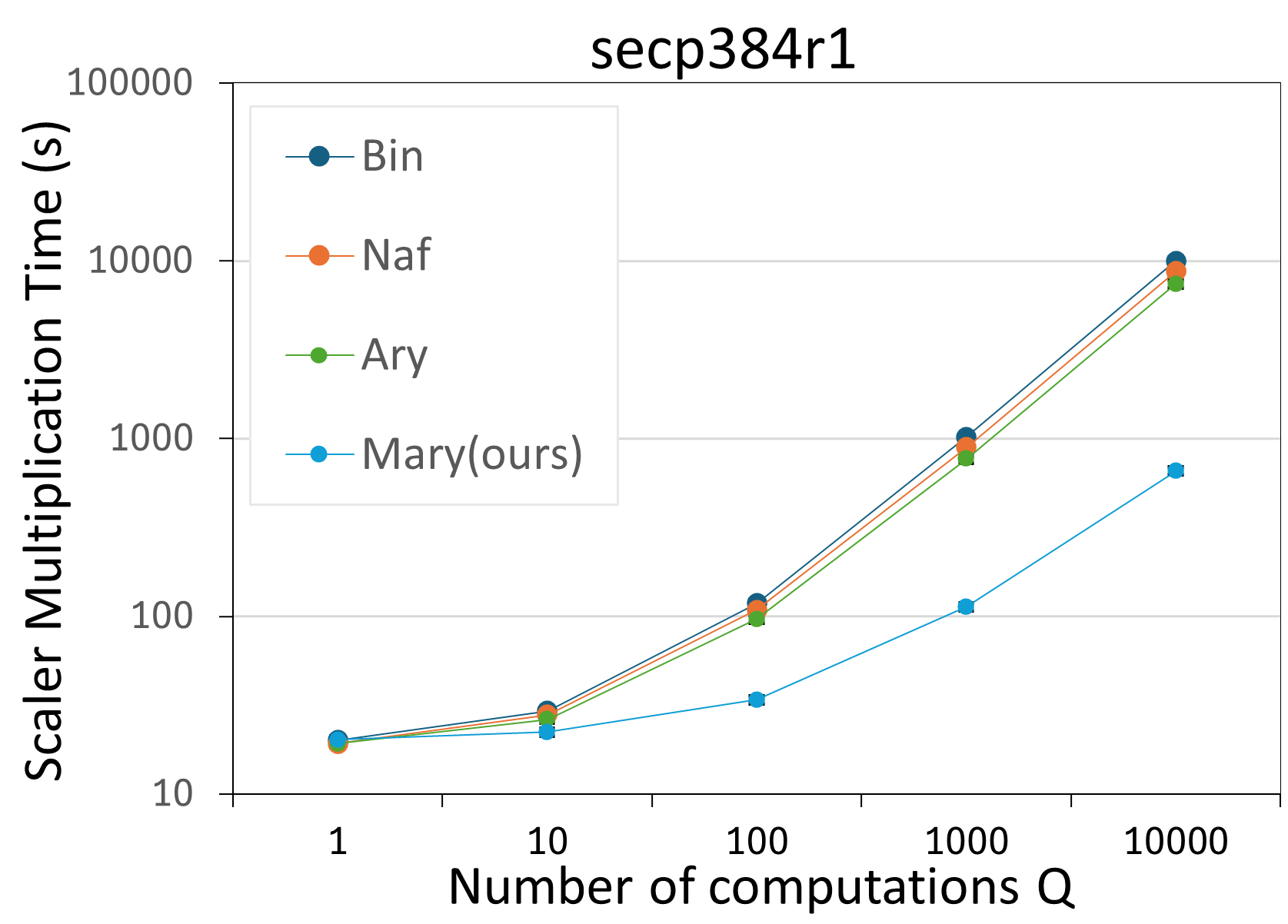}%
\label{fig:exp1b}}
\hfil
\subfloat[]{\includegraphics[width=0.3\textwidth]{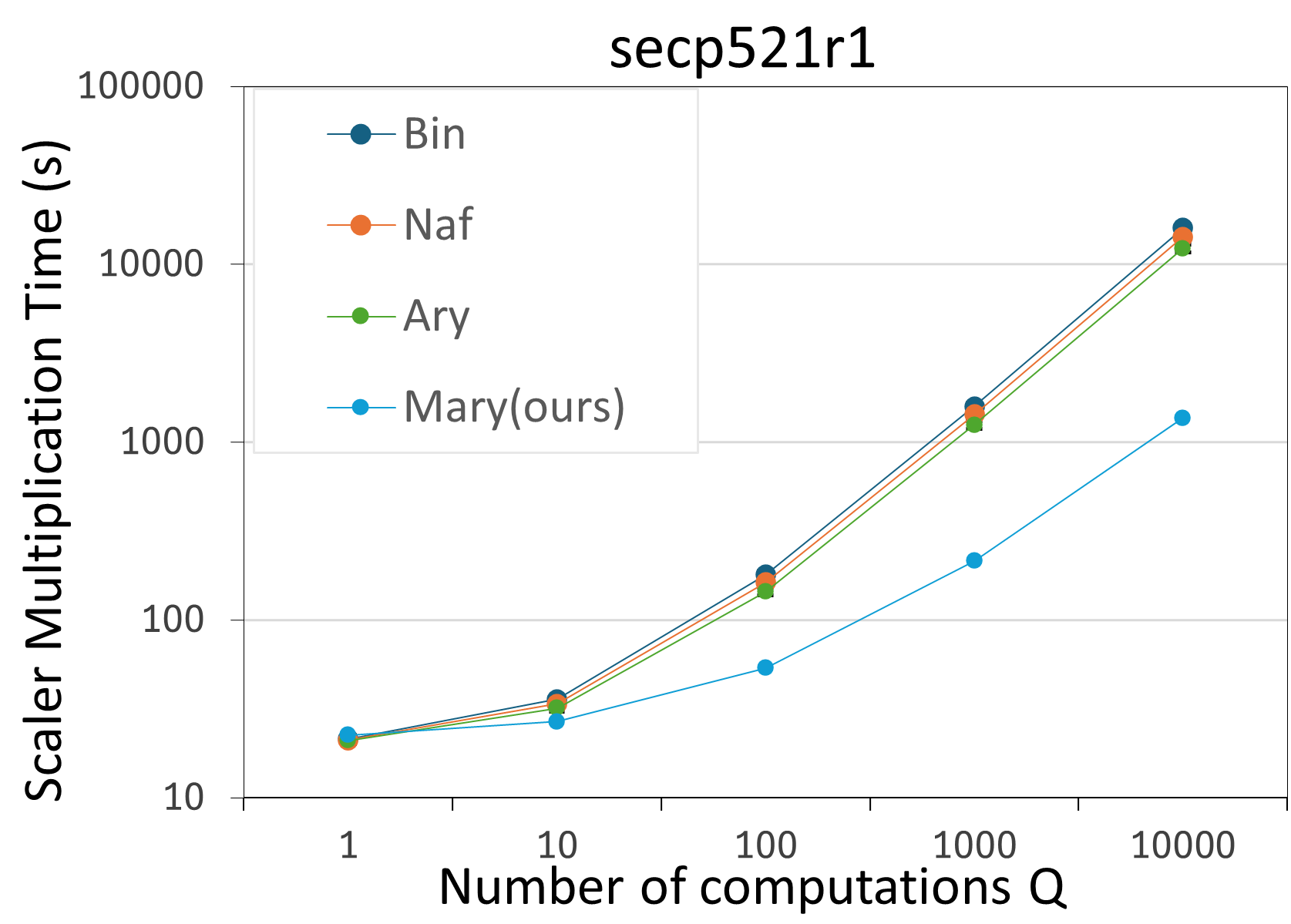}%
\label{fig:exp1c}}
\captionsetup{justification=centering,singlelinecheck=false}
\caption{Evaluation on the time consumed by various optimized algorithms for scalar multiplication with increasing number of computations $ Q $. Specifically, \Cref{fig:exp1a}, \Cref{fig:exp1b}, and \Cref{fig:exp1c} represent the runtime for the elliptic curves secp256k1, secp384r1, and secp521r1, respectively.}
\label{fig:exp1}
\end{figure*}

This evaluation investigates the impact of the number of computations $ Q $ and base size $ p $ on the efficiency of four scalar multiplication algorithms for points on the elliptic curve $ E_p(a, b) $. Quantitative results are presented in \Cref{fig:exp1}, where Figures \Cref{fig:exp1a}, \Cref{fig:exp1b}, and \Cref{fig:exp1c} illustrate the comparative efficiency across different base field sizes $ p $.

\begin{figure*}[!t]
\centering
\subfloat[]{\includegraphics[width=0.3\textwidth]{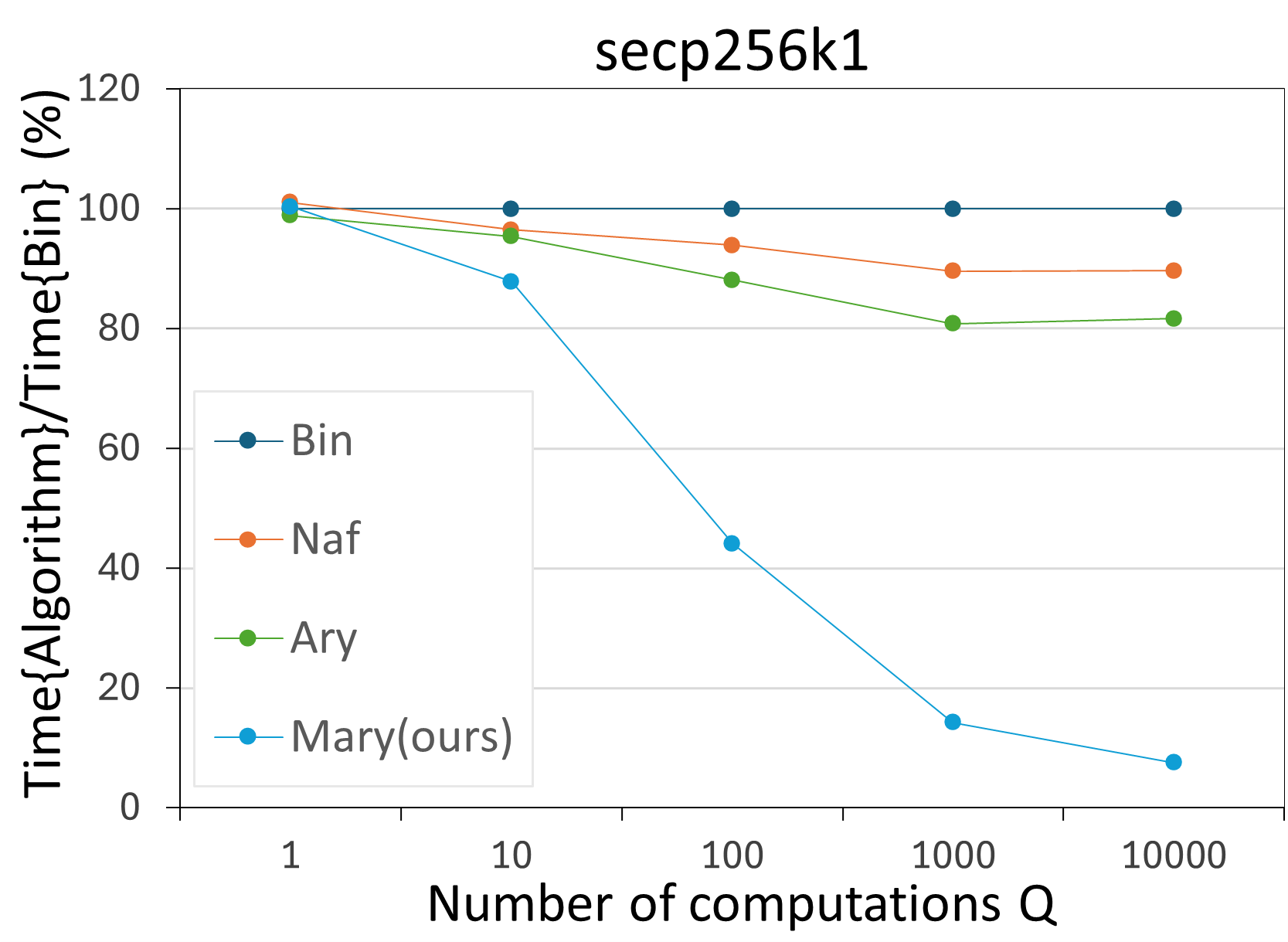}%
\label{fig:exp3a}}
\hfil
\subfloat[]{\includegraphics[width=0.3\textwidth]{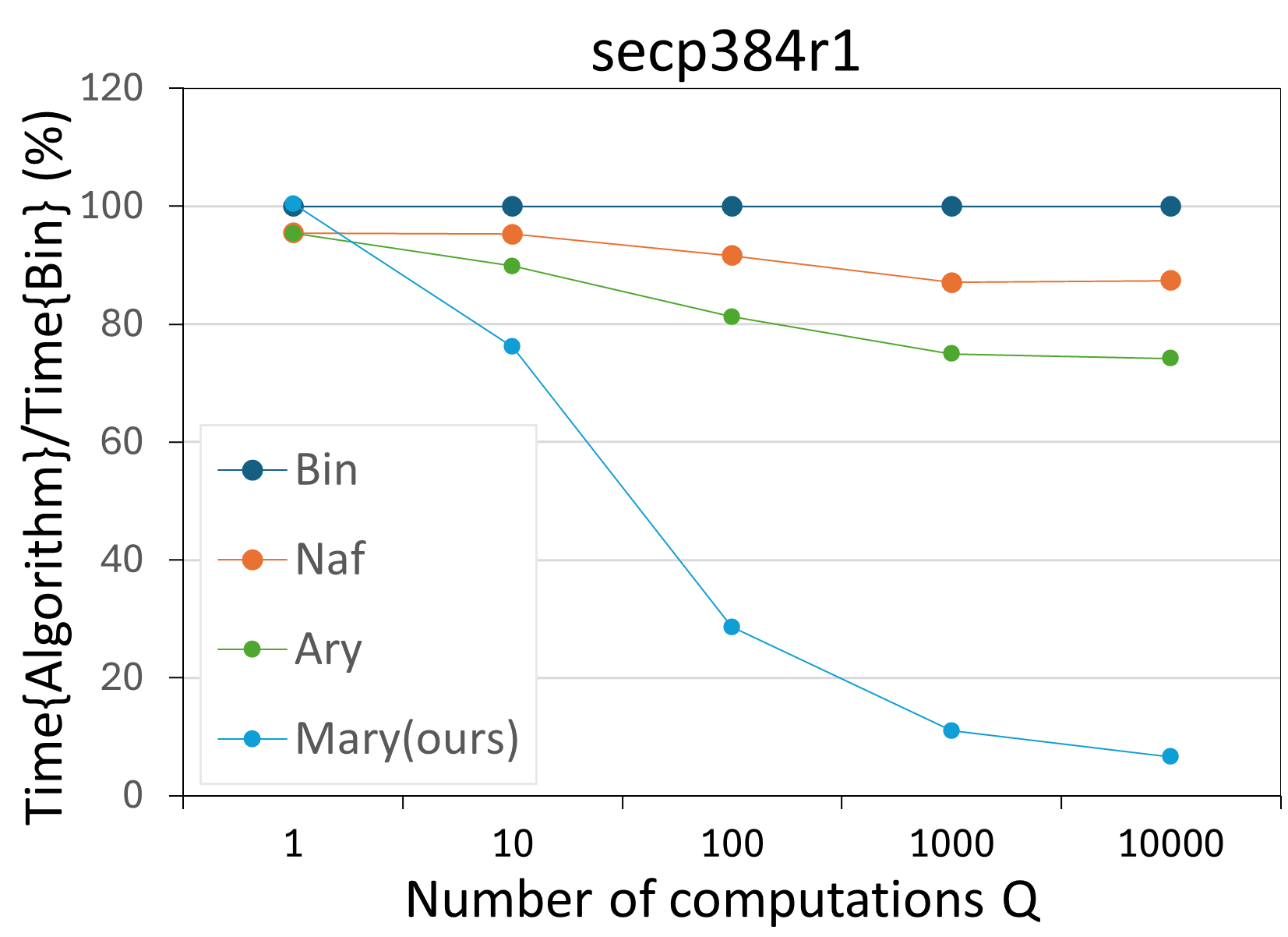}%
\label{fig:exp3b}}
\hfil
\subfloat[]{\includegraphics[width=0.3\textwidth]{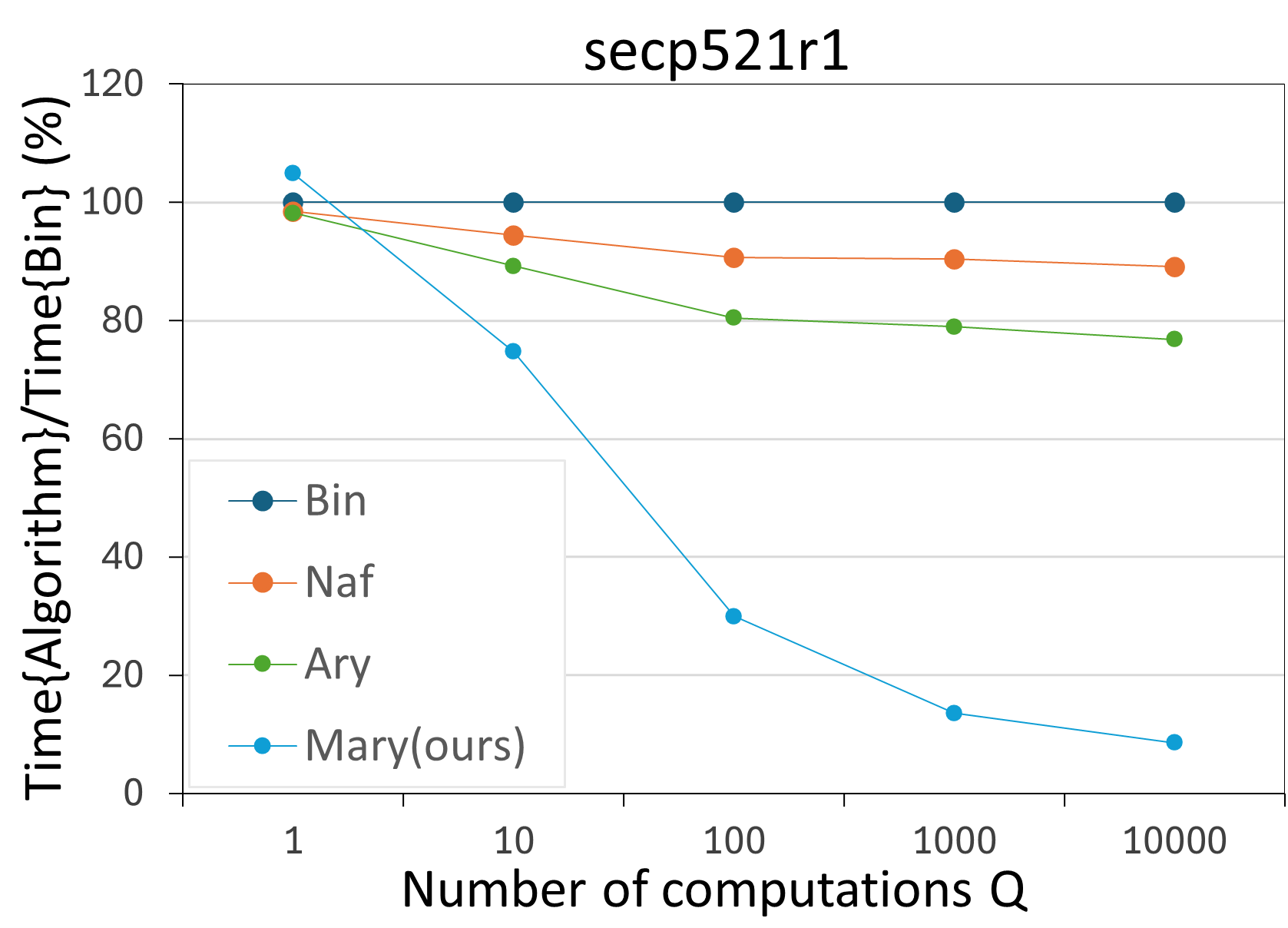}%
\label{fig:exp3c}}
\captionsetup{justification=centering,singlelinecheck=false}
\caption{Evaluation on the proportion of time consumed by various optimized algorithms for scalar multiplication relative to the double-and-add algorithm as the number of computations $ Q $ increases. Specifically, \Cref{fig:exp3a}, \Cref{fig:exp3b}, and \Cref{fig:exp3c} represent the proportion for the elliptic curves secp256k1, secp384r1, and secp521r1, respectively.}
\label{fig:exp3}
\end{figure*}

As $ Q $ increases, the efficiency advantage of our proposed algorithm over other optimized algorithms becomes more pronounced. \Cref{fig:exp3} shows that when $ Q $ is large, our algorithm consumes approximately 10\% of the time required by the Double-and-Add algorithm, while the NAF-based algorithm takes about 90\% and the $ 2^k $-ary algorithm about 79\%. This improvement is due to the time complexity of precomputation-based optimization algorithms, which includes a factor of $ \frac{1}{\log Q} $; hence, as $ Q $ increases, the optimization effect intensifies. These experimental results align with our theoretical analysis.

\begin{figure*}[!t]
\captionsetup{justification=centering, singlelinecheck=false}
\centering
\subfloat[]{\includegraphics[width=0.3\textwidth]{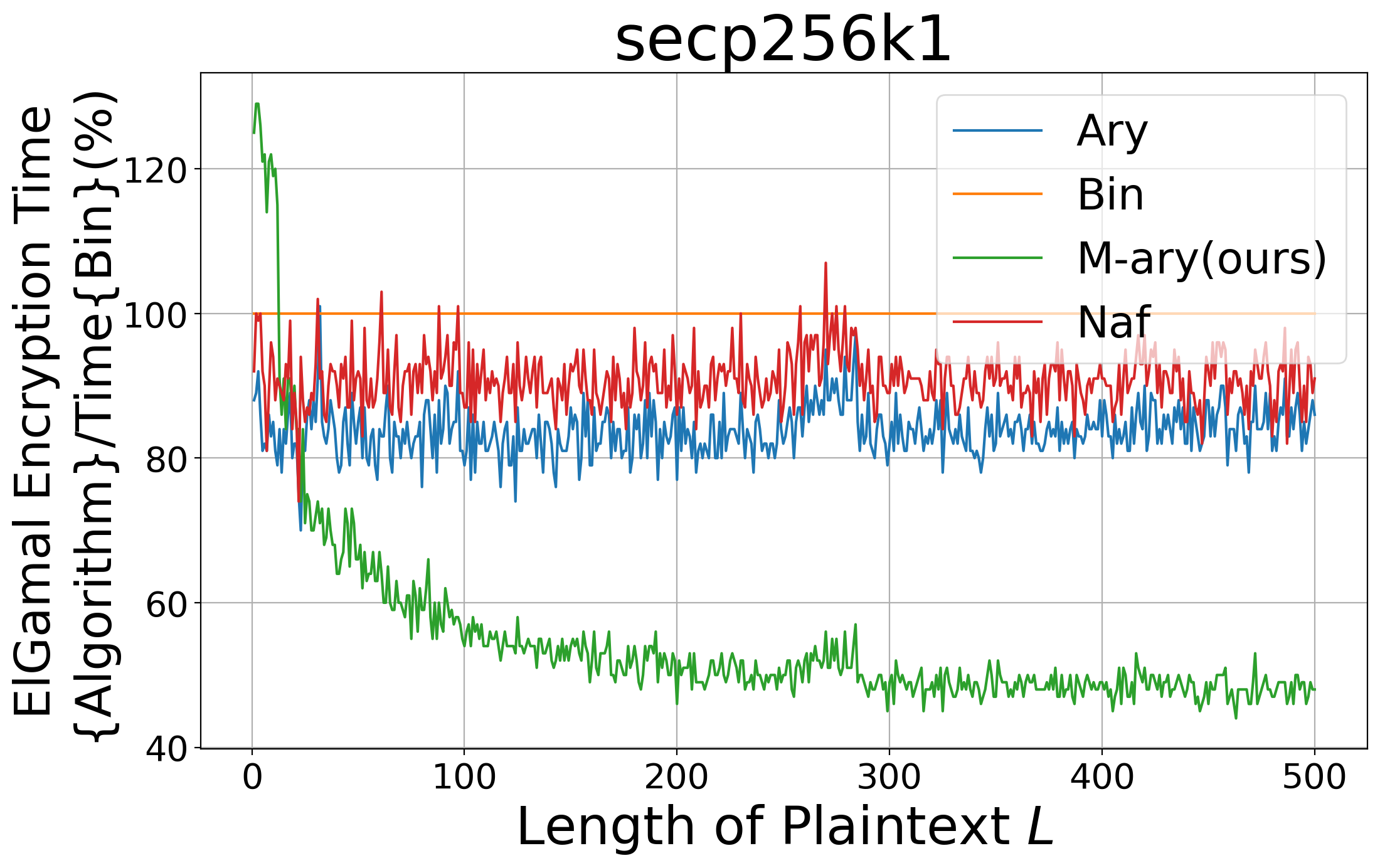}%
\label{fig:plaintext1}}
\hfil
\subfloat[]{\includegraphics[width=0.3\textwidth]{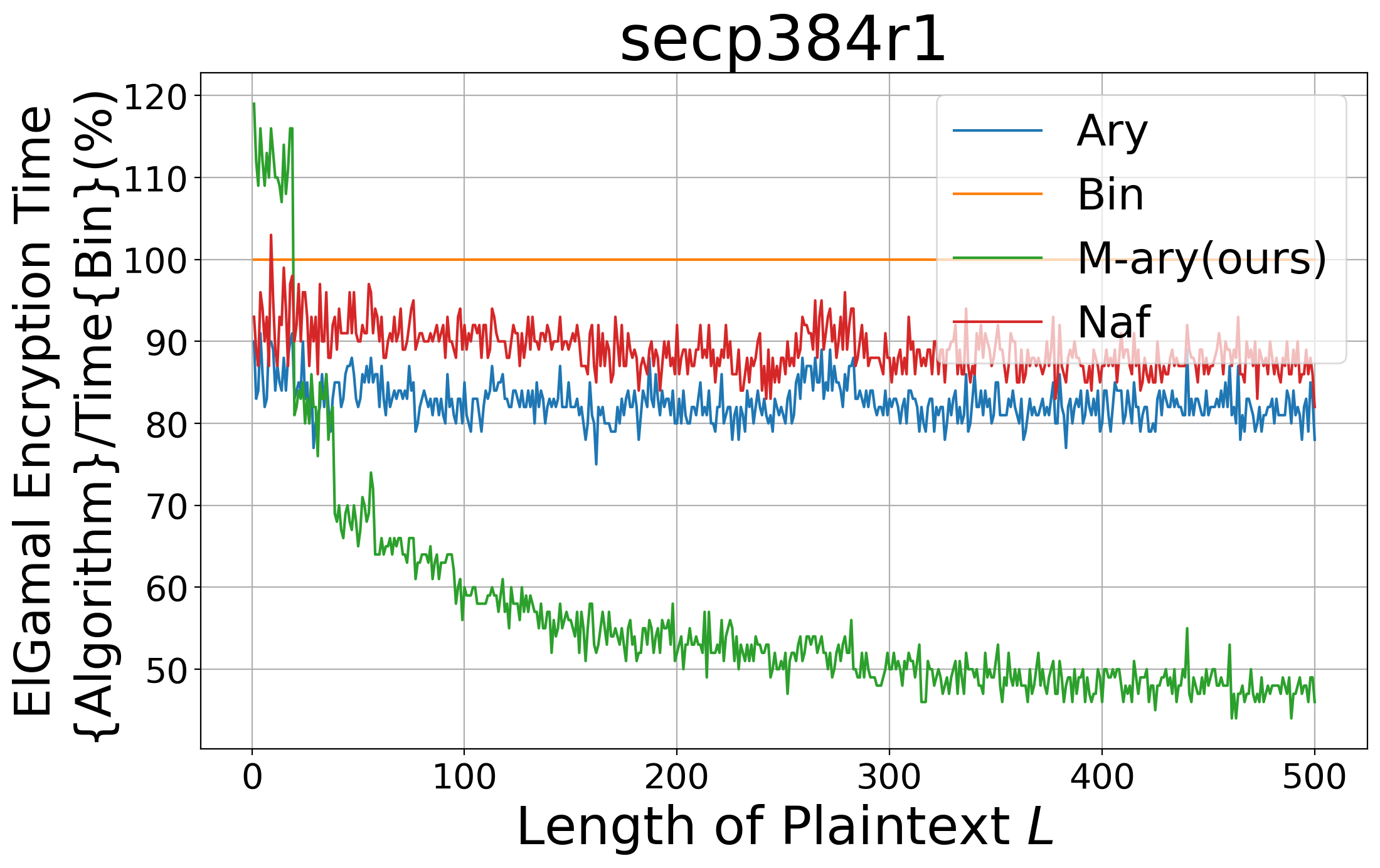}%
\label{fig:plaintext2}}
\hfil
\subfloat[]{\includegraphics[width=0.3\textwidth]{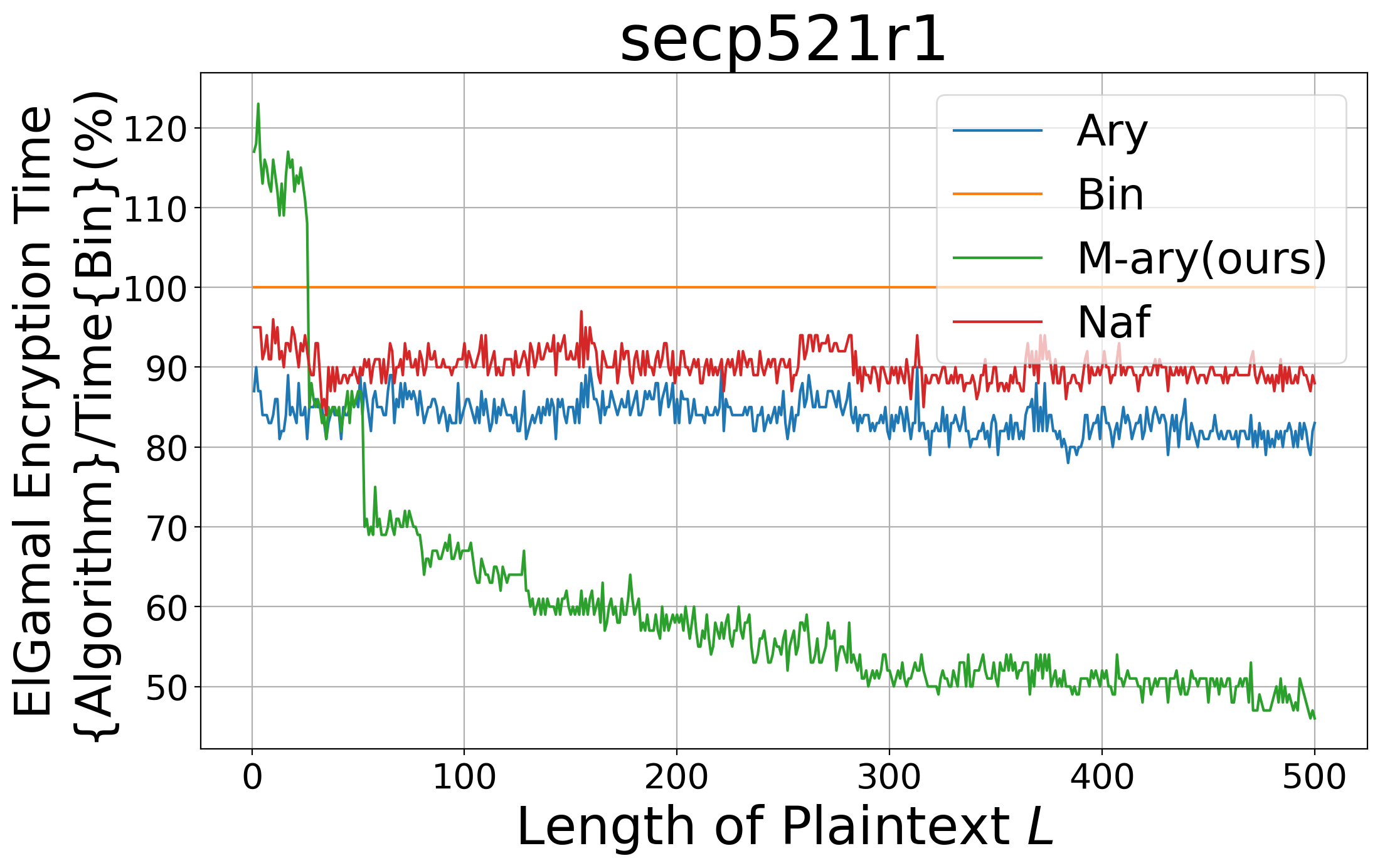}%
\label{fig:plaintext3}}
\captionsetup{justification=centering, singlelinecheck=false}
\caption{Evaluation of the proportion of time consumed by ElGamal Encryption using various optimized scalar multiplication algorithms compared to the double-and-add algorithm, as the plaintext length \( L \) increases. Specifically, \Cref{fig:plaintext1}, \Cref{fig:plaintext2}, and \Cref{fig:plaintext3} represent the respective proportions for the elliptic curves secp256k1, secp384r1, and secp521r1.}
\label{fig:plaintext}
\end{figure*}


\subsubsection{Evaluation on the Performance of Different Scalar Multiplication Optimization Algorithms in ElGamal Cryptosystem}
We evaluate our M-ary Precomputation-Based Accelerated Scalar Multiplication Algorithm within the ElGamal cryptosystem, comparing its performance against traditional methods such as Double-and-Add, NAF-based, and \(2^k\)-ary algorithms. 

During encryption, User B computes \(C_1 = kG\) and \(C_2 = P_m + kP_A\), transmitting the ciphertext \(C_m = (C_1, C_2)\) to User A. Since the base point \(G\) and public key \(P_A\) are fixed, our M-ary algorithm accelerates scalar multiplications for \(kG\) and \(kP_A\). The algorithm's advantages increase with the number of scalar multiplications \(Q\).

To map a message \(m\) to a point on the elliptic curve \(E\), we use a probabilistic mapping algorithm that encodes \(m\) into \(P_m\). Each message is divided into groups of size \(k\) and represented as base-256 numbers mapped to points on \(E\). Our algorithm computes all scalar multiplications simultaneously, significantly reducing encryption time.

\begin{table}[!t]
    \caption{Comparative Analysis of Time Consumed by Different Scalar Multiplication Algorithms used in ElGamal encryption when Length of Plaintext $ L=100,300,360$.\label{tbl:time}}
    \centering
    \scriptsize
    \begin{tabular}{cccccc}
    \toprule[1.5pt]
    \multirow{2}{*}{Algorithm} & \multirow{2}{*}{Curve} & \multicolumn{3}{c}{Time Consumed (seconds)} \\
    \cmidrule{3-5}
     & & $ L=100 $ & $ L=300 $ & $ L=360 $ \\
    \midrule[1pt]
    \multirow{3}{*}{Double-and-add} 
     & secp256k1 & 0.18s & 0.49s & 0.79s \\
     & secp384r1 & 0.26s & 0.71s & 1.16s \\
     & secp521r1 & 0.36s & 1.10s & 1.83s \\
    \midrule[1pt]
    \multirow{3}{*}{NAF-based} 
     & secp256k1 & 0.16s & 0.44s & 0.72s \\
     & secp384r1 & 0.24s & 0.62s & 0.95s \\
     & secp521r1 & 0.33s & 0.96s & 1.61s \\
    \midrule[1pt]
    \multirow{3}{*}{$2^{k}$-ary} 
     & secp256k1 & 0.14s & 0.40s & 0.68s \\
     & secp384r1 & 0.22s & 0.59s & 0.91s \\
     & secp521r1 & 0.30s & 0.89s & 1.51s \\
    \midrule[1pt]
    \multirow{3}{*}{\textbf{M-ary Precomputation (ours)}} 
     & secp256k1 & \textbf{0.10s} & \textbf{0.24s} & \textbf{0.38s} \\
     & secp384r1 & \textbf{0.15s} & \textbf{0.35s} & \textbf{0.54s} \\
     & secp521r1 & \textbf{0.24s} & \textbf{0.57s} & \textbf{0.85s} \\
    \toprule[1.5pt]
\end{tabular}
\end{table}

Using elliptic curves secp256k1, secp384r1, and secp521r1, we compared the time for ElGamal encryption across various plaintext lengths \(L\). As shown in \Cref{fig:plaintext} and \Cref{tbl:time}, our method outperforms traditional algorithms as \(L\) increases. Specifically, when \(L\) is large, the NAF and \(2^k\)-ary algorithms require 78\% to 92\% of the time of the Double-and-Add method, while our approach only requires about 46\%.


These results demonstrate the superior performance of our algorithm, particularly as the scalar size \(L\) increases. This improvement stems from the time complexity factor \( \frac{1}{\log Q} \) in our method, which ensures enhanced efficiency as \(Q\) scales with \(L\). The empirical findings validate the theoretical analysis and highlight the strong scalability of our approach, making it well-suited for high-performance cryptographic applications. These results provide a solid foundation for further evaluation in broader communication and security contexts.

\subsection{Evaluation in Communication-Oriented Simulation Using NS3 Framework}

To further assess the practical performance of the proposed scalar multiplication algorithms, we conducted a series of simulations using the NS3 framework. These experiments were designed to evaluate the algorithms within a representative communication network setting, focusing on key metrics such as encryption time, communication time, and overall simulation time.

\begin{figure*}[!ht]
\centering
\captionsetup{justification=centering, singlelinecheck=false}
\subfloat[\scriptsize Communication Network Topology: Demonstrating point-to-point connections and the associated communication parameters such as data rates and delays.]{
\includegraphics[width=2.6in]{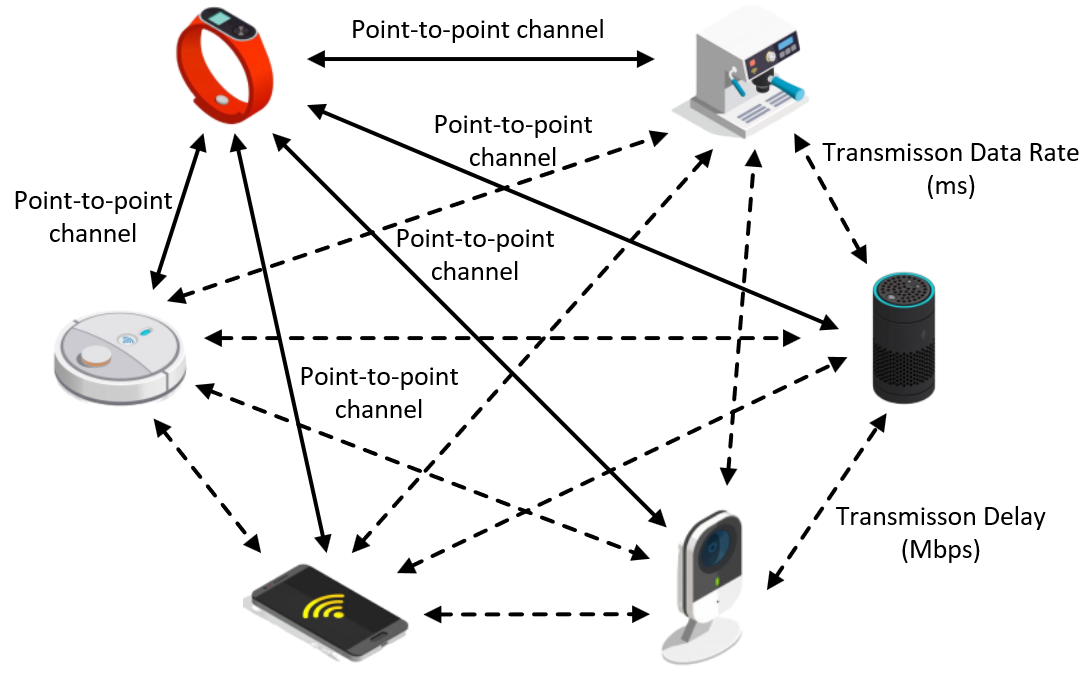}%
\label{fig:iot1}}
\hfil
\subfloat[\scriptsize NS3 Simulation Network Architecture: Illustrating the node components, protocol stack, and point-to-point communication channels.]{
\includegraphics[width=2.6in]{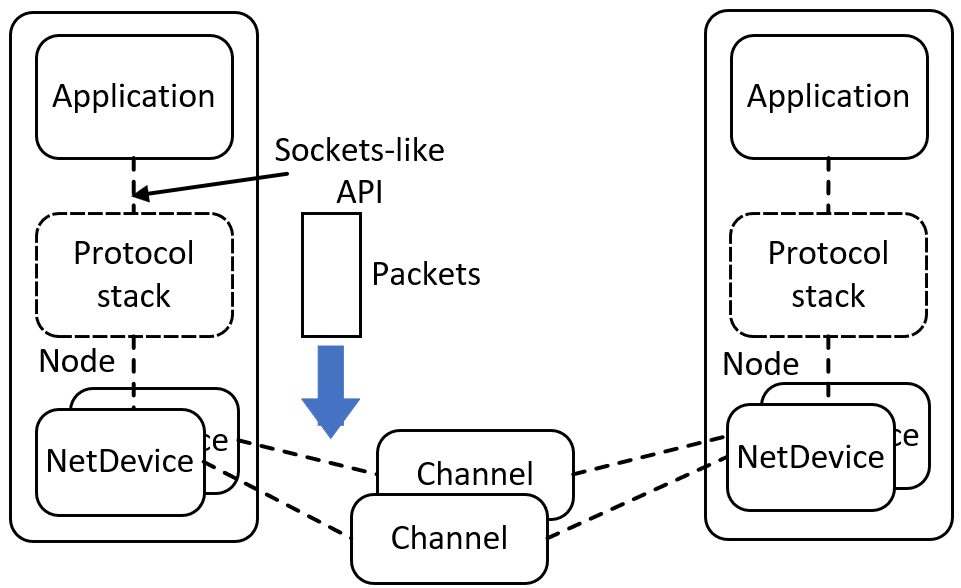}%
\label{fig:iot2}}
\caption{
\scriptsize NS3 simulation communication model. \Cref{fig:iot1} demonstrates the overall network topology with point-to-point links and their associated communication metrics, such as transmission delays and data rates. \Cref{fig:iot2} illustrates the internal architecture of nodes in the NS3 simulation, showing the application layer, protocol stack, and net device components.
}
\end{figure*}

\subsubsection{Simulation Setup}
The simulation environment modeled a communication network with five interconnected nodes, each representing a device performing secure data transmission. As illustrated in \Cref{fig:iot1}, the nodes were connected using point-to-point channels with a bandwidth of 5 Mbps and a delay of 2 ms, reflecting typical parameters for lightweight communication systems.

To evaluate the performance of different scalar multiplication algorithms, we encrypted a set of predefined text messages of varying sizes, simulating practical data transmission scenarios. The elliptic curve-based ElGamal encryption scheme was employed, incorporating eight scalar multiplication algorithms: Double-and-Add, NAF-based, \(2^k\)-ary, Sliding Window, Montgomery Ladder, Fixed-Base Comb, Fixed-Window, and our proposed M-ary precomputation-based algorithm, including both the standard and binary-optimized versions.

Three primary performance indicators were recorded during the simulations:
\begin{itemize}
    \item \textbf{Total Encryption Time (Enc. Time)}: The total time required to encrypt all messages using the selected scalar multiplication algorithm.
    \item \textbf{Total Communication Time (Com. Time)}: The overall duration required to transmit all encrypted messages across the network.
    \item \textbf{Total Simulation Time (Sim. Time)}: The end-to-end duration of the complete simulation process, including encryption, transmission, and protocol handling.
\end{itemize}

Messages were encrypted prior to transmission using the selected algorithm, and the corresponding timing results were logged for comprehensive evaluation. Each node's internal structure, depicted in \Cref{fig:iot2}, comprises an application layer, a protocol stack, and network devices, reflecting a typical layered communication architecture.

\subsubsection{Simulation Results}

\begin{table*}[!htbp]
\centering
\caption{Simulation Results for Different Scalar Multiplication Algorithms and Elliptic Curves.}
\scriptsize
\label{tab:sim_results}
\begin{tabular}{@{}p{2.1cm}p{3cm}p{2.5cm}p{2.5cm}p{2.5cm}@{}}
\toprule
\textbf{Elliptic Curve} & \textbf{Algorithm} & \textbf{Enc. Time (s)} & \textbf{Com. Time (s)} & \textbf{Sim. Time (s)} \\ \midrule
\multirow{6}{*}{secp256k1} 
& Double-and-add          & 6.8971    & 6.8978    & 6.9041    \\ \cline{2-5} 
& NAF-based               & 6.8802    & 6.8810    & 6.8909    \\ \cline{2-5} 
& $2^k$-ary              & 6.8032    & 6.8039    & 6.8115    \\ \cline{2-5} 
& Sliding Window          & 6.6316    & 6.6324    & 6.6397    \\ \cline{2-5} 
& Montgomery Ladder       & 7.1261    & 7.1268    & 7.1360    \\ \cline{2-5} 
& Fixed-Base Comb       & 6.7629    & 6.7636    & 6.7729    \\ \cline{2-5} 
& Fixed-Window       & 6.8547     & 6.8554    & 6.8618    \\ \cline{2-5} 
& \textbf{M-ary (ours)}   & \textbf{6.3361} & \textbf{6.3369} & \textbf{6.3491} \\
\cline{2-5}
& \textbf{M-ary (binary)}   & 6.61878 & 6.6197 & 6.6279 \\ \midrule
\multirow{6}{*}{secp384r1} 
& Double-and-add          & 8.5375    & 8.5382    & 8.5444    \\ \cline{2-5} 
& NAF-based               & 8.3399    & 8.3407    & 8.3483    \\ \cline{2-5} 
& $2^k$-ary               & 7.7314    & 7.7321    & 7.7385    \\ \cline{2-5} 
& Sliding Window          & 7.9080    & 7.9087    & 7.9150    \\ \cline{2-5} 
& Montgomery Ladder       & 9.2238    & 9.2245    & 9.2338    \\ \cline{2-5} 
& Fixed-Base Comb       & 7.4359    & 7.4366    & 7.4431    \\ \cline{2-5} 
& Fixed-Window       & 8.2275    & 8.2282    & 8.2345    \\ \cline{2-5} 
& \textbf{M-ary (ours)}   & \textbf{7.0521} & \textbf{7.0528} & \textbf{7.0602} \\ 
\cline{2-5}
& \textbf{M-ary (binary)}   & 7.2438 & 7.2445 & 7.25407 \\ \midrule
\multirow{6}{*}{secp521r1} 
& Double-and-add          & 10.9273    & 10.9281    & 10.9376    \\ \cline{2-5} 
& NAF-based               & 10.6670    & 10.6679    & 10.7002    \\ \cline{2-5} 
& $2^k$-ary               & 9.7595     & 9.7603     & 9.7699    \\ \cline{2-5} 
& Sliding Window          & 9.6087     & 9.6094      & 9.6160    \\ \cline{2-5} 
& Montgomery Ladder       & 12.2517   & 12.2525   & 12.2626   \\ \cline{2-5} 
& Fixed-Base Comb       & 11.8115     & 11.8123   & 11.8227   \\ \cline{2-5} 
& Fixed-Window       & 10.4312    & 10.4319    & 10.4416    \\ \cline{2-5} 
& \textbf{M-ary (ours)}   & \textbf{8.3926} & \textbf{8.3934} & \textbf{8.4002} \\
\cline{2-5}
& \textbf{M-ary (binary)}   & 8.5293 & 8.5300 & 8.5404 \\
\bottomrule
\end{tabular}
\end{table*}

\noindent{\textbf{Time Efficiency Analysis.}}  
The NS3 simulation results demonstrate the enhanced efficiency of our proposed M-ary precomputation-based algorithm across different elliptic curve settings. Table~\ref{tab:sim_results} summarizes the encryption time, communication time, and overall simulation time for each algorithm across three elliptic curves: secp256k1, secp384r1, and secp521r1. Specifically, our method consistently achieves lower encryption times compared to other algorithms, leading to reductions in overall communication and simulation times, and highlighting its scalability for large-scale cryptographic applications.

\begin{figure*}[!t]
\captionsetup{justification=centering, singlelinecheck=false}
\centering
\subfloat[]{
\includegraphics[width=0.3\textwidth]{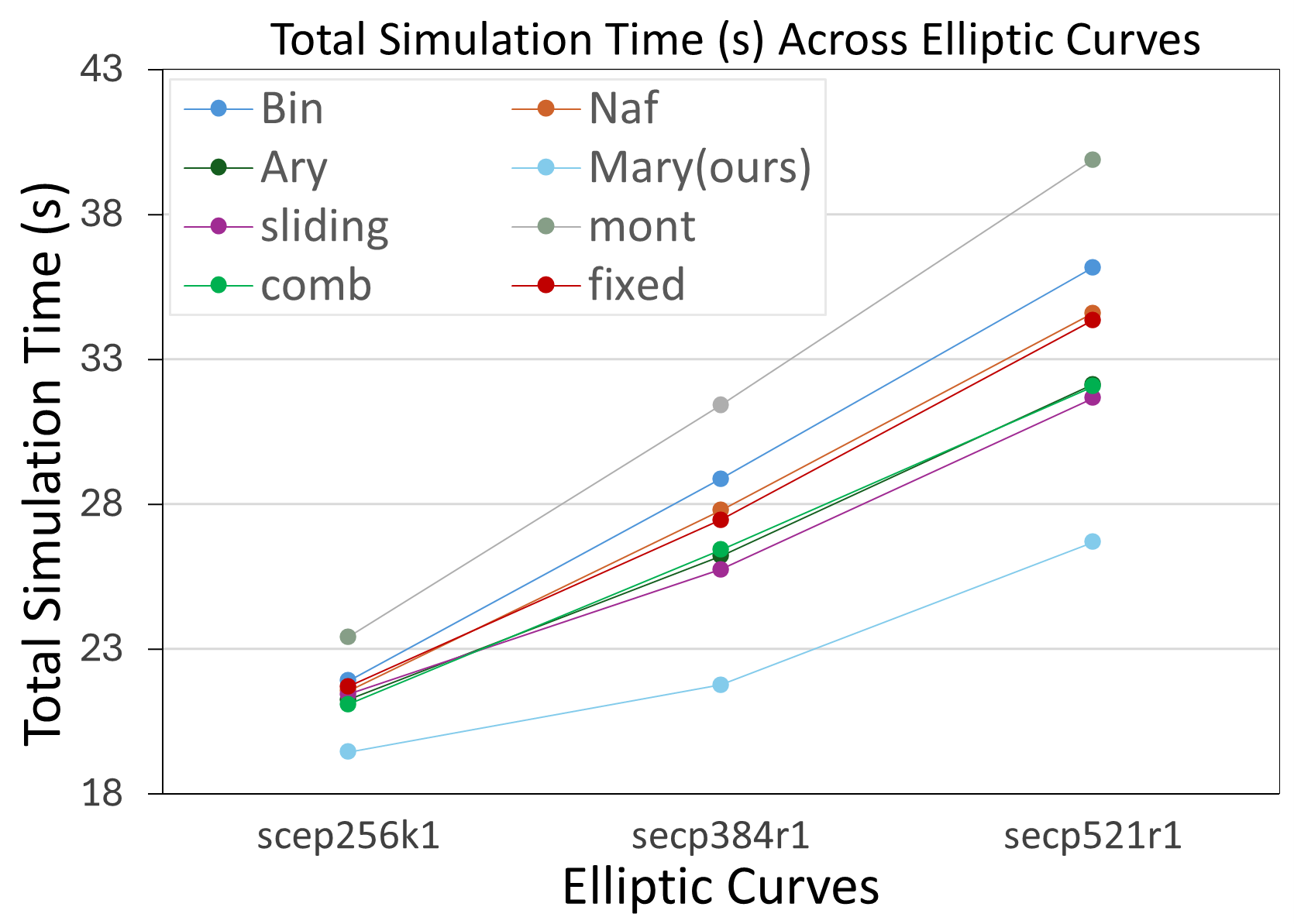}%
\label{fig:exp2a}}
\hfil
\subfloat[]{
\includegraphics[width=0.3\textwidth]{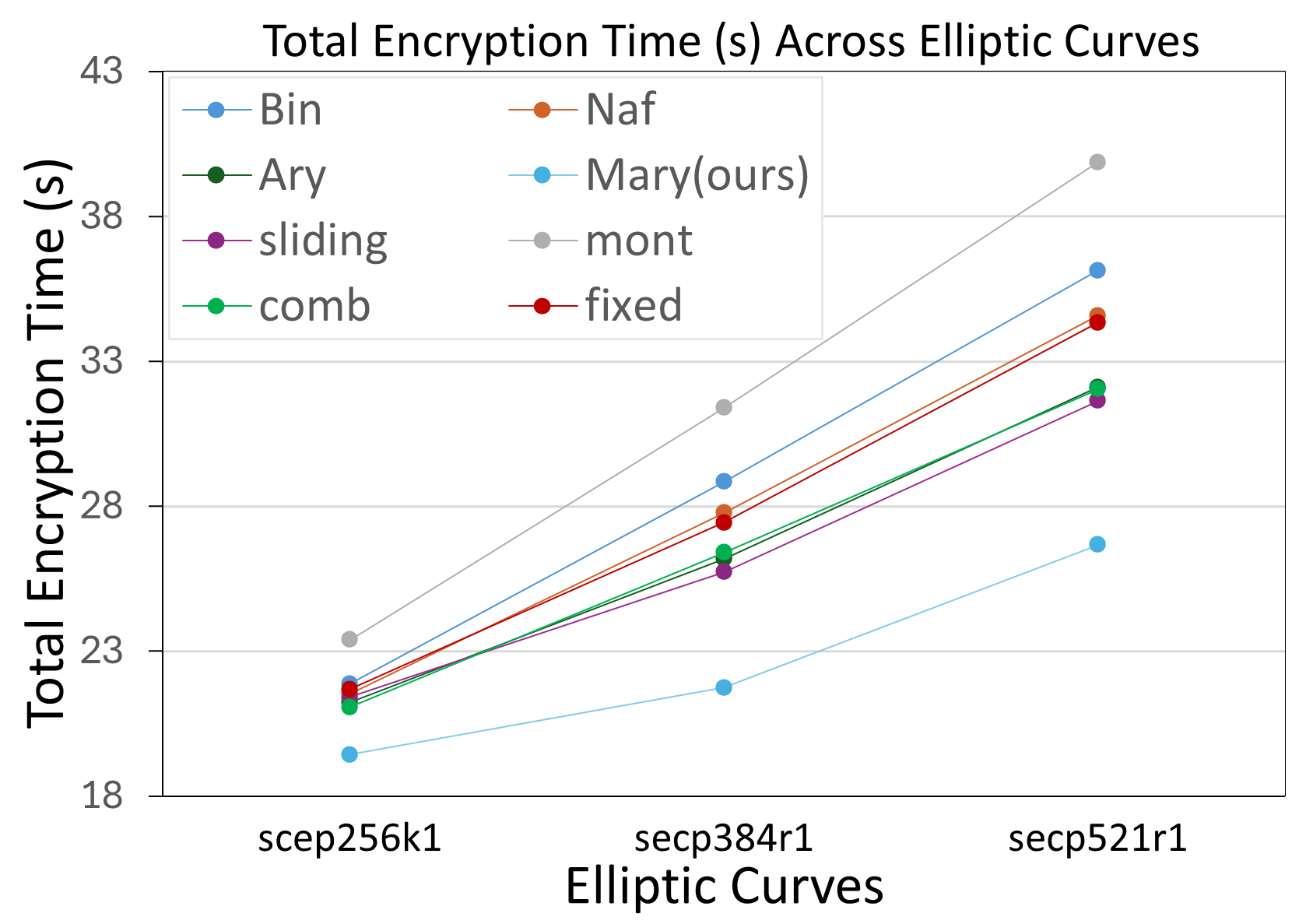}%
\label{fig:exp2b}}
\hfil
\subfloat[]{
\includegraphics[width=0.3\textwidth]{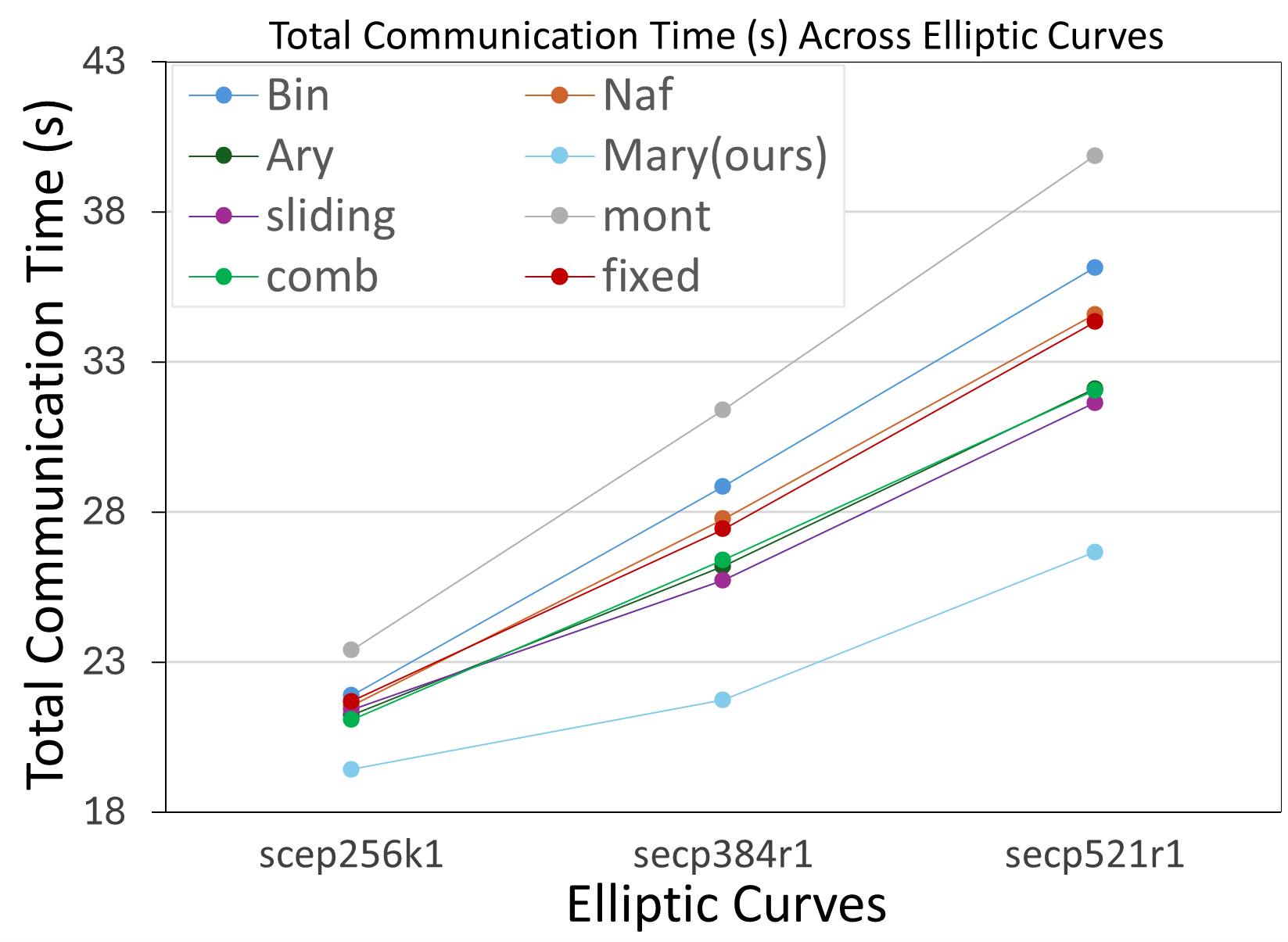}%
\label{fig:exp2c}}
\caption{
Performance Comparison of Scalar Multiplication Algorithms on Elliptic Curves. \Cref{fig:exp2a} shows the total simulation time across three elliptic curves (secp256k1, secp384r1, secp521r1). \Cref{fig:exp2b} depicts the total encryption time. \Cref{fig:exp2c} presents the total communication time.
}
\end{figure*}

Figures~\ref{fig:exp2a}, \ref{fig:exp2b}, and \ref{fig:exp2c} provide a visual comparison of the performance metrics:
\begin{itemize}
    \item Figure~\ref{fig:exp2a} shows the reduction in encryption time across multiple transmission sessions, highlighting the computational efficiency of the proposed method.
    \item Figure~\ref{fig:exp2b} depicts communication time savings, demonstrating the improved transmission performance achieved by our approach.
    \item Figure~\ref{fig:exp2c} illustrates that our algorithm consistently outperforms others in terms of total simulation time across all tested elliptic curves, showcasing its strong scalability with increasing computational complexity.
\end{itemize}

These results underscore the superior performance and practicality of our M-ary precomputation-based algorithm, making it highly suitable for high-performance and large-scale cryptographic applications that require both speed and computational efficiency.

\noindent{\textbf{Memory Consumption Analysis.}} 
Peak memory usage is one of the most critical metrics for evaluating the deployability and scalability of scalar multiplication algorithms. Rather than only focusing on the theoretical size of precomputation tables, we measure the \textit{peak memory usage} during the entire execution process, including temporary variables, intermediate buffers, and any auxiliary structures used during computation. This better reflects the algorithm’s actual memory footprint in practical systems.

To systematically compare the memory performance of different algorithms, we selected three standard elliptic curves—secp256k1, secp384r1, and secp521r1—and measured the peak memory usage for each of nine scalar multiplication algorithms under four scalar counts $Q \in \{1, 10, 100, 1000\}$. All measurements reflect the actual peak memory usage during execution, not just theoretical memory requirements. A comprehensive summary of the experimental results is provided in Table~\ref{tab:memory_usage}.

\begin{table*}[htbp]
\centering
\caption{Peak memory usage (in MB) of different scalar multiplication algorithms across three curves and four scalar counts $Q \in \{1, 10, 100, 1000\}$. Values in bold represent the two algorithms achieving the lowest peak memory consumption in each configuration.}
\label{tab:memory_usage}
\tiny
\begin{tabular}{llcccc}
\toprule
Elliptic Curve & Algorithm & $Q=1$ & $Q=10$ & $Q=100$ & $Q=1000$ \\
\midrule
\multirow{9}{*}{\textbf{secp256k1}} 
& Double-and-add     & 0.000969 & 0.002602 & 0.018410 & 0.177040 \\
& NAF-based          & 0.000931 & 0.002560 & 0.018368 & 0.177002 \\
& $2^k$-ary          & 0.003941 & 0.005093 & 0.016094 & 0.126659 \\
& Sliding Window     & 0.032118 & 0.033269 & 0.044271 & 0.165089 \\
& Montgomery Ladder  & 0.001081 & \textbf{0.002232} & \textbf{0.013234} & \textbf{0.123797} \\
& Fixed-Base Comb    & 0.001528 & 0.003160 & 0.018969 & 0.161898 \\
& Fixed-Window       & 0.007359 & 0.008991 & 0.020153 & 0.130718 \\
& \textbf{M-ary (ours)}       & \textbf{0.000797} & 0.002430 & 0.018238 & 0.144852 \\
& \textbf{M-ary (binary)}     & \textbf{0.000690} & \textbf{0.001842} & \textbf{0.012844} & \textbf{0.123409} \\
\midrule
\multirow{9}{*}{\textbf{secp384r1}} 
& Double-and-add     & 0.001152 & 0.003059 & 0.021614 & 0.207710 \\
& NAF-based          & 0.001129 & 0.003036 & 0.021591 & 0.207687 \\
& $2^k$-ary          & 0.004871 & 0.006298 & 0.020046 & 0.158077 \\
& Sliding Window     & 0.040190 & 0.041616 & 0.055364 & 0.201939 \\
& Montgomery Ladder  & 0.001370 & \textbf{0.002797} & \textbf{0.016544} & \textbf{0.154574} \\
& Fixed-Base Comb    & 0.001787 & 0.003695 & 0.022248 & 0.192642 \\
& Fixed-Window       & 0.009449 & 0.011356 & 0.025265 & 0.163296 \\
& \textbf{M-ary (ours)}       & \textbf{0.000950} & 0.002857 & 0.021412 & 0.186707 \\
& \textbf{M-ary (binary)}     & \textbf{0.000843} & \textbf{0.002270} & \textbf{0.016018} & \textbf{0.154049} \\
\midrule
\multirow{9}{*}{\textbf{secp521r1}} 
& Double-and-add     & 0.001377 & 0.003628 & 0.025616 & 0.246044 \\
& NAF-based          & 0.001389 & 0.003635 & 0.025627 & 0.246052 \\
& $2^k$-ary          & 0.005772 & 0.007542 & 0.024723 & 0.197086 \\
& Sliding Window     & 0.050177 & 0.051948 & 0.069129 & 0.248326 \\
& Montgomery Ladder  & 0.001706 & 0.003477 & \textbf{0.020658} & \textbf{0.193020} \\
& Fixed-Base Comb    & 0.002146 & 0.004439 & 0.026430 & 0.231155 \\
& Fixed-Window       & 0.011677 & 0.014477 & 0.031818 & 0.203709 \\
& \textbf{M-ary (ours)}       & \textbf{0.001152} & \textbf{0.003403} & 0.025391 & 0.236874 \\
& \textbf{M-ary (binary)}     & \textbf{0.001041} & \textbf{0.002815} & \textbf{0.019997} & \textbf{0.192360} \\
\bottomrule
\end{tabular}
\end{table*}

Traditional algorithms such as Double-and-Add and NAF-based exhibit moderate memory usage overall. While they do not involve precomputation, their peak memory is not necessarily minimal. For instance, on secp384r1 with $Q = 100$, NAF-based consumes 0.0216 MB—higher than both Montgomery Ladder (0.0165 MB) and our M-ary (binary) method (0.0160 MB).

Fixed-Base Comb and Fixed-Window algorithms tend to show higher memory consumption, especially in small-scale tasks. For example, on secp256k1 with $Q = 10$, Fixed-Window uses 0.0090 MB, whereas our M-ary (binary) method requires only 0.0018 MB—almost 5 times smaller.

The Sliding Window algorithm, despite being configured with the minimal window size in our experiments, still incurs significant memory overhead due to its exponential precomputation structure. On secp521r1 with $Q = 1$, it reaches a peak of 0.0502 MB, while M-ary (binary) only uses 0.0010 MB—demonstrating a more than 50-fold reduction.

Montgomery Ladder, while not a precomputation-based method, achieves consistently low memory usage due to its simple structure without extra caching. However, it still falls short of our optimized binary method in most settings. Our original M-ary algorithm demonstrates stable performance across tasks, achieving lower memory usage than most traditional methods while maintaining computational efficiency.

\textbf{Our M-ary (binary) algorithm consistently achieves the best memory efficiency}. It ranks first in 9 out of 12 test cases, and top-two in all remaining cases. For example, on secp256k1 with $Q = 10$, it uses just 0.0018 MB—lower than all other compared methods. On secp384r1 with $Q = 100$, it consumes only 0.0160 MB (the lowest among all). On secp521r1, it ranks among the lowest in every $Q$ setting. This optimization benefits from a sparse power representation, storing only $2^j$ multiples in each layer and reconstructing the target value through dynamic additions, thereby significantly reducing memory redundancy.

\section{Potential for Hardware Implementation}

Although this paper mainly focuses on theoretical analysis, the proposed M-ary precomputation-based scalar multiplication algorithm shows strong potential for hardware acceleration on platforms such as GPUs and FPGAs.

\begin{figure*}[ht]
\captionsetup{justification=centering, singlelinecheck=false}
  \centering
  \subfloat[\scriptsize gECC Framework for GPU-based Batch ECC Acceleration]{
    \includegraphics[width=0.60\textwidth]{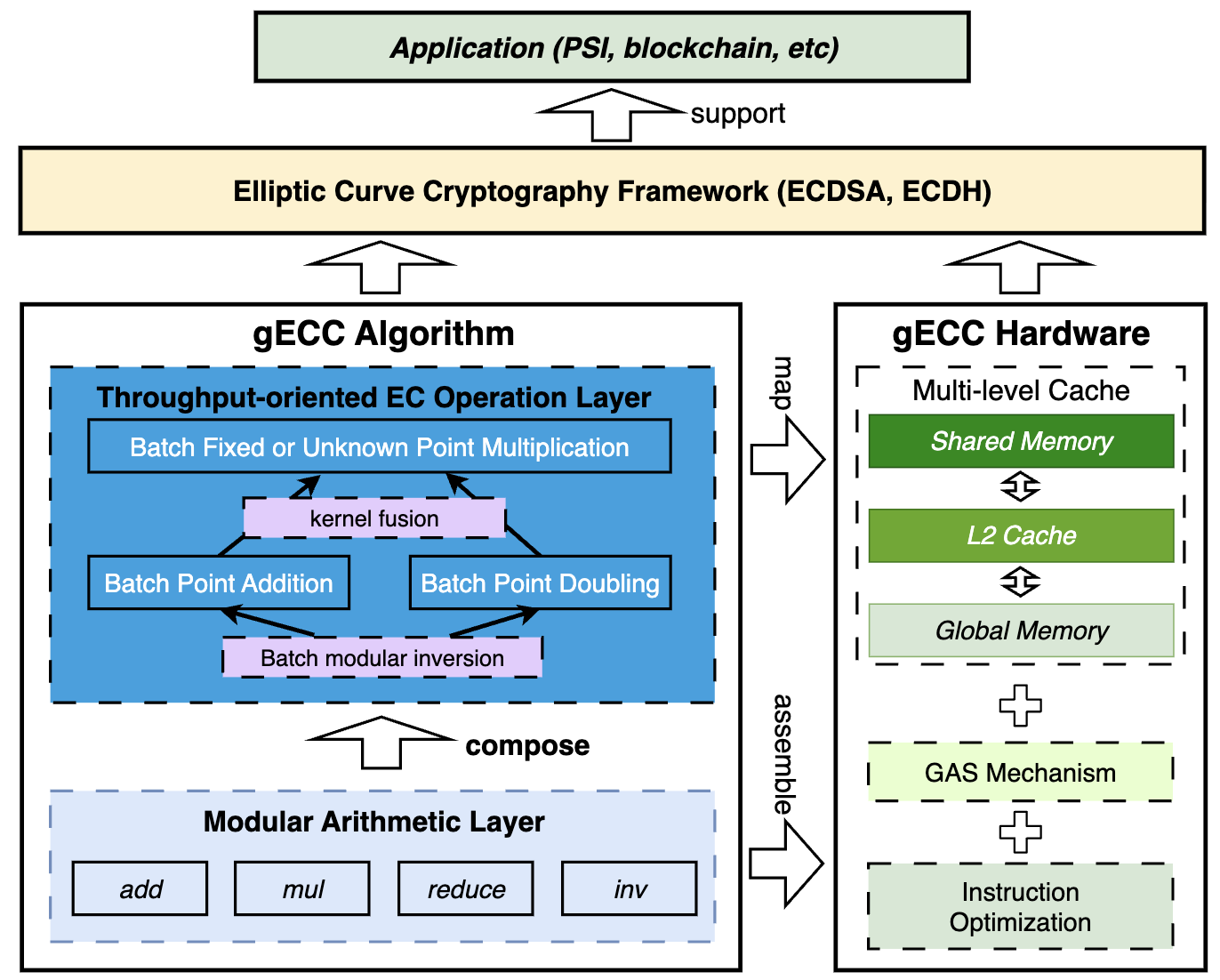}
    \label{fig:gpu}
  }
  \hfil
  \subfloat[\scriptsize FPGA Architecture for Koblitz Curve Point Multiplication]{
    \includegraphics[width=0.80\textwidth]{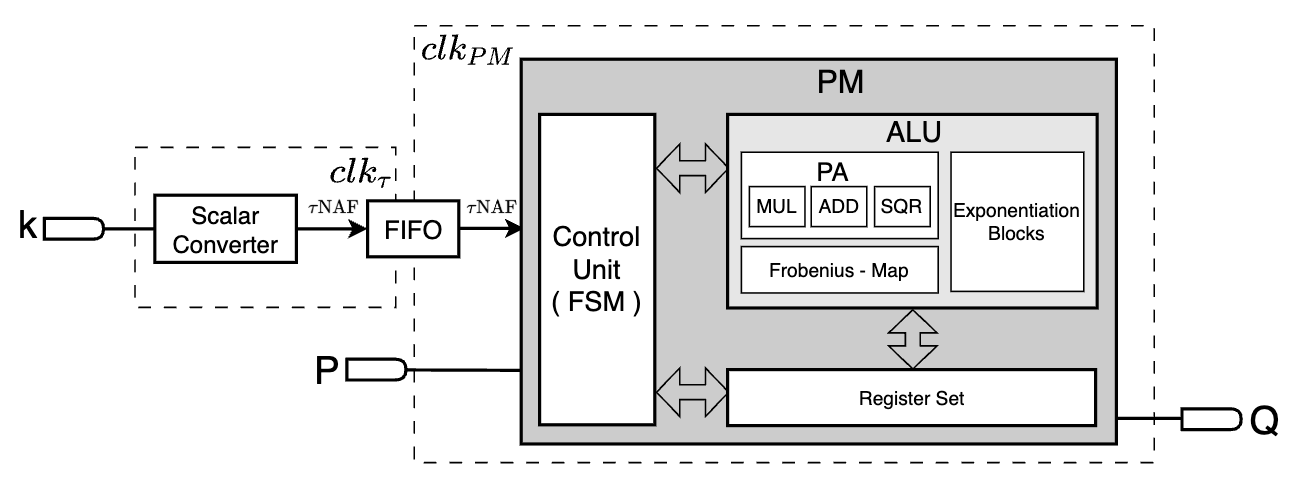}
    \label{fig:fpga}
  }
  \caption{Hardware Deployment Potentials of the proposed M-ary scalar multiplication method: (a) GPU-based high-throughput ECC framework; (b) FPGA-based scalar conversion and point multiplication architecture.}
  \label{fig:hardware_potential}
\end{figure*}

Recent studies, such as the gECC framework proposed by Xiong et al.~\cite{xiong2024gecc}, have demonstrated that optimizing batch scalar multiplications on GPUs can significantly improve the throughput of elliptic curve cryptographic (ECC) operations. Their framework, illustrated in \Cref{fig:gpu}, leverages Montgomery’s trick and efficient scheduling strategies to reduce memory access overhead, achieving high performance on NVIDIA A100 GPUs. Given the reduced time and memory complexity of our M-ary method, it is well-suited for integration into such parallel architectures, promising further improvements in latency and throughput.

On FPGA platforms, Marzouqi et al.~\cite{marzouqi2015high} presented a high-speed ECC processor based on redundant signed digit (RSD) representations and pipelined Karatsuba–Ofman multipliers. Their results highlight that optimized scalar decompositions can effectively reduce scalar multiplication latency and area, which aligns closely with the objectives of our M-ary method.

Moreover, Jiang et al.~\cite{jiang2025low} proposed low-latency and area-efficient point multiplication architectures over Koblitz curves. Their overall hardware architecture, shown in \Cref{fig:fpga}, optimizes scalar conversion and computation scheduling within pipelined designs to achieve significant latency reductions. Although their work specifically targets Koblitz curves, the core design principles—precomputation strategies, pipelined scheduling, and area-time trade-off optimization—are broadly applicable to general ECC scalar multiplication problems.

Therefore, the proposed M-ary algorithm not only improves theoretical efficiency but also exhibits strong potential for hardware compatibility. Future work will explore implementing the algorithm on modern GPU and FPGA platforms to validate its practical performance gains.

\section{Conclusion}
This paper presents an M-ary precomputation-based accelerated scalar multiplication algorithm for elliptic curve cryptography (ECC), designed to enhance computational efficiency and optimize memory usage. By leveraging structured precomputation and scalar decomposition, the proposed method reduces the time complexity from \( \Theta(Q \log p) \) to \( \Theta\left(\frac{Q \log p}{\log Q}\right) \) and achieves a flexible memory complexity of \( \Theta\left(\frac{Q \log p}{\log^2 Q}\right) \), enabling efficient operation across a wide range of computational scales.

Comprehensive evaluations validate the effectiveness of the proposed approach. Theoretical analysis confirms its superior asymptotic efficiency compared to traditional scalar multiplication methods. In cryptographic experiments using the ElGamal encryption scheme, the proposed method achieves up to a 59\% reduction in encryption time on the secp256k1 curve when \(Q=1000\), compared to Double-and-Add, NAF-based, and \(2^k\)-ary algorithms. In NS3-based simulations, the M-ary (binary) variant achieves a 22\% reduction in total communication time on the secp384r1 curve and up to a 25\% reduction in overall simulation time on the secp521r1 curve. Memory performance is also significantly improved: for instance, on secp256k1 with \(Q=1000\), the M-ary (binary) method reduces peak memory usage by approximately 25\% compared to the Sliding Window algorithm.

These results collectively underscore the versatility, scalability, and practical applicability of the M-ary precomputation-based scalar multiplication algorithm. The method not only enhances cryptographic performance but also provides a promising foundation for deployment in secure communication systems, large-scale cryptographic computations, and hardware-accelerated cryptographic applications.

\section{Future Work}

This work primarily focuses on optimizing scalar multiplication in single-scalar scenarios with a fixed base point. Several promising directions for future research are identified:

\begin{itemize}
    \item \textbf{Extension to Multi-Scalar Multiplication (MSM):}  
    Extending the proposed M-ary precomputation method to multi-scalar multiplication, which plays a critical role in applications such as pairing-based cryptography and signature aggregation, is a natural progression. Investigating dynamic parameter selection strategies for efficient multi-scalar operations remains an open challenge.

    \item \textbf{Parallelization and Batch Processing:}  
    Exploring parallel and batch execution of the algorithm could further enhance its computational efficiency, particularly in settings where large volumes of scalar multiplications must be processed simultaneously.

    \item \textbf{Adaptation to Dynamic Curve Parameters and Base Points:}  
    While this work assumes fixed elliptic curve parameters and base points, many real-world cryptographic systems involve variable curves or changing base points. Future research may explore incremental table update mechanisms to efficiently adapt precomputation structures without incurring significant recomputation overhead.

    \item \textbf{Hardware-Optimized Implementations:}  
    Investigating optimized deployments of the algorithm on diverse hardware architectures, such as ARM, RISC-V, and GPU platforms, could further demonstrate its practical scalability and performance across different computational environments.
\end{itemize}

These directions offer valuable opportunities to further enhance the applicability, efficiency, and robustness of the proposed M-ary precomputation-based scalar multiplication method across a broad range of cryptographic applications.

\section*{Acknowledgments}
This work is supported by the National Natural Science Foundation of China (No. 62162057, No. 61872254), the Key Lab of Information Network Security of the Ministry of Public Security, China (C20606), the Sichuan Science and Technology Program, China (2021JDRC0004), and the Key Laboratory of Data Protection and Intelligent Management of the Ministry of Education, China (SCUSAKFKT202402Y). We express our gratitude to the corresponding authors, Tongxi Wu and Xufeng Liu, for their equal contributions and invaluable advice throughout the writing process of this paper.

\appendix
\section{Parameters of the Selected Elliptic Curves for Evaluation}
We selected three elliptic curves defined by the Standards for Efficient Cryptography Group (SECG) for evaluation: secp256k1, secp384r1, and secp521r1. Each elliptic curve is defined by a quintuple \(T = (p, a, b, G, n)\). The parameters are as follows (all represented in hexadecimal):

\subsection{secp256k1}

The finite field \(\mathbb{F}_p\) of secp256k1 is defined by the prime \(p\):
\begin{align*}
p &= \text{FFFFFFFF FFFFFFFF FFFFFFFF} \\
  &\quad\text{FFFFFFFF FFFFFFFF FFFFFFFF} \\
  &\quad\text{FFFFFFFF FFFFFC2F} \\
  &= 2^{256} - 2^{32} - 2^9 - 2^8 - 2^7 - 2^6 - 2^4 - 1
\end{align*}

The elliptic curve \(E\) over \(\mathbb{F}_p\) is defined by the Weierstrass equation:
\[
E: y^2 = x^3 + ax + b,
\]
where 
\begin{align*}
a &= \text{00000000 00000000 00000000} \\
  &\quad\text{00000000 00000000 00000000} \\
  &\quad\text{00000000 00000000} \\
b &= \text{00000000 00000000 00000000} \\
  &\quad\text{00000000 00000000 00000000} \\
  &\quad\text{00000000 00000007} \\
\end{align*}

The base point \(G = (x, y)\) on the curve is defined as:
\begin{align*}
x &= \text{79BE667E F9DCBBAC 55A06295} \\
  &\quad\text{CE870B07 029BFCDB 2DCE28D9} \\
  &\quad\text{59F2815B 16F81798} \\
y &= \text{483ADA77 26A3C465 5DA4FBFC} \\
  &\quad\text{0E1108A8 FD17B448 A6855419} \\
  &\quad\text{9C47D08F FB10D4B8} \\
\end{align*}

The order \(n\) of the base point \(G\) is:
\begin{align*}
n &= \text{FFFFFFFF FFFFFFFF FFFFFFFF} \\
  &\quad\text{FFFFFFFF BAAEDCE6 AF48A03B} \\
  &\quad\text{BFD25E8C D0364141} \\
\end{align*}

\subsection{secp384r1}
The finite field \(\mathbb{F}_p\) of secp384r1 is defined as:
\begin{align*}
p &= \text{FFFFFFFF FFFFFFFF FFFFFFFF} \\
  &\quad\text{FFFFFFFF FFFFFFFF FFFFFFFF} \\
  &\quad\text{00000000 00000000 FFFFFFFF} \\
  &= 2^{384} - 2^{128} - 2^{96} + 2^{32} - 1
\end{align*}

The elliptic curve \(E\) over \(\mathbb{F}_p\) is defined by the Weierstrass equation \(y^2 = x^3 + ax + b\), where \(a\) and \(b\) are:
\begin{align*}
a &= \text{FFFFFFFF FFFFFFFF FFFFFFFF} \\
  &\quad\text{FFFFFFFE FFFFFFFF FFFFFFFF} \\
  &\quad\text{FFFFFFFE FFFFFFFE FFFFFFFF} \\
  &\quad\text{00000000 00000000 FFFFFFFC} \\
b &= \text{B3312FA7 E23EE7E4 988E056B} \\
  &\quad\text{E3F82D19 181D9C6E FE814112} \\
  &\quad\text{0314088F 5013875A C656398D} \\
  &\quad\text{8A2ED19D 2A85C8ED D3EC2AEF} \\
\end{align*}

The base point \(G = (x, y)\) has coordinates:
\begin{align*}
x &= \text{AA87CA22 BE8B0537 8EB1C71E} \\
  &\quad\text{F320AD74 6E1D3B62 8BA79B98} \\
  &\quad\text{59F741E0 82542A38 5502F25D} \\
  &\quad\text{BF55296C 3A545E38 72760AB7} \\
y &= \text{3617DE4A 96262C6F 5D9E98BF} \\
  &\quad\text{9292DC29 F8F41DBD 289A147C} \\
  &\quad\text{E9DA3113 B5F0B8C0 0A60B1CE} \\
  &\quad\text{1D7E819D 7A431D7C 90EA0E5F} \\
\end{align*}

The order \(n\) of the base point \(G\) is:
\begin{align*}
n &= \text{FFFFFFFF FFFFFFFF FFFFFFFF} \\
  &\quad\text{FFFFFFFF FFFFFFFF FFFFFFFF} \\
  &\quad\text{C7634D81 F4372DDF 581A0DB2} \\
  &\quad\text{48B0A77A ECEC196A CCC52973} \\
\end{align*}

\subsection{secp521r1}
The finite field \(\mathbb{F}_p\) of secp521r1 is defined as:
\begin{align*}
p &= \text{01FF FFFFFFFF FFFFFFFF} \\
  &\quad\text{FFFFFFFF FFFFFFFF FFFFFFFF} \\
  &\quad\text{FFFFFFFF FFFFFFFF FFFFFFFF} \\
  &\quad\text{FFFFFFFF FFFFFFFF FFFFFFFF} \\
  &\quad\text{FFFFFFFF FFFFFFFF FFFFFFFF} \\
  &\quad\text{FFFFFFFF FFFFFFFF} \\
  &= 2^{521} - 1
\end{align*}

The elliptic curve \(E\) over \(\mathbb{F}_p\) is defined by the Weierstrass equation \(y^2 = x^3 + ax + b\), where \(a\) and \(b\) are:
\begin{align*}
a &= \text{01FF FFFFFFFF FFFFFFFF} \\
  &\quad\text{FFFFFFFF FFFFFFFF FFFFFFFF} \\
  &\quad\text{FFFFFFFF FFFFFFFF FFFFFFFF} \\
  &\quad\text{FFFFFFFF FFFFFFFF FFFFFFFF} \\
  &\quad\text{FFFFFFFF FFFFFFFF FFFFFFFF} \\
  &\quad\text{FFFFFFFF FFFFFFFC} \\
b &= \text{0051 953EB961 8E1C9A1F} \\
  &\quad\text{929A21A0 B68540EE A2DA725B} \\
  &\quad\text{99B315F3 B8B48991 8EF109E1} \\
  &\quad\text{56193951 EC7E937B 1652C0BD} \\
  &\quad\text{3BB1BF07 3573DF88 3D2C34F1} \\
  &\quad\text{EF451FD4 6B503F00} \\
\end{align*}

The base point \(G = (x, y)\) has coordinates:
\begin{align*}
x &= \text{00C6 858E06B7 0404E9CD} \\
  &\quad\text{9E3ECB66 2395B442 9C648139} \\
  &\quad\text{053FB521 F828AF60 6B4D3DBA} \\
  &\quad\text{A14B5E77 EFE75928 FE1DC127} \\
  &\quad\text{A2FFA8DE 3348B3C1 856A429B} \\
  &\quad\text{F97E7E31 C2E5BD66} \\
y &= \text{0118 39296A78 9A3BC004} \\
  &\quad\text{5C8A5FB4 2C7D1BD9 98F54449} \\
  &\quad\text{579B4468 17AFBD17 273E662C} \\
  &\quad\text{97EE7299 5EF42640 C550B901} \\
  &\quad\text{3FAD0761 353C7086 A272C240} \\
  &\quad\text{88BE9476 9FD16650} \\
\end{align*}

The order \(n\) of the base point \(G\) is:
\begin{align*}
n &= \text{01FF FFFFFFFF FFFFFFFF} \\
  &\quad\text{FFFFFFFF FFFFFFFF FFFFFFFF} \\
  &\quad\text{FFFFFFFF FFFFFFFF FFFFFFFA} \\
  &\quad\text{51868783 BF2F966B 7FCC0148} \\
  &\quad\text{F709A5D0 3BB5C9B8 899C47AE} \\
  &\quad\text{BB6FB71E 91386409} \\
\end{align*}

\section{Mathematical Proofs and Theorems}

\subsection{Proof of NAF Representation and Its Properties}

This section presents theorems and proofs foundational to the NAF-based scalar multiplication algorithm.

\begin{list}{}{}
    \item {\textbf{Theorem 1.} An integer \( n \) can be represented in signed binary form as a sequence \( a_{d-1}, a_{d-2}, \ldots, a_0 \), satisfying:
    \begin{equation}
        n = \sum_{i=0}^{d-1} a_i \cdot 2^i,
    \end{equation}
    where \( a_i \in \{-1, 0, 1\} \) and \( a_{d-1} \neq 0 \).}
    
    \item {\textbf{Theorem 2.} The number of non-zero elements in the signed binary representation of \( n \) is the Hamming weight, denoted as:
    \begin{equation}
        \sum_{i=0}^{d-1} [a_i \neq 0].
    \end{equation}}
    
    \item {\textbf{Theorem 3.} The signed binary representation of \( n \) is called the Non-Adjacent Form (NAF) if \( a_i a_{i+1} = 0 \) for all \( 0 \leq i < d-1 \). It is denoted as \( (a_{d-1} a_{d-2} \ldots a_1 a_0)_{\text{NAF}} \).}
    
    \item {\textbf{Theorem 4.} The NAF of a non-negative integer \( n \) is unique and denoted as \( \mathrm{NAF}(n) \).}
    
    \item {\textbf{Theorem 5.} The Hamming weight of \( \mathrm{NAF}(n) \) is the smallest among all signed binary representations of \( n \).}
    
    \item {\textbf{Theorem 6.} The length \( \ell(n) \) of the NAF of a positive integer \( n \) satisfies:
    \begin{equation}
        \left\lfloor \log_2 n \right\rfloor + 1 \leq \ell(n) \leq \left\lceil \log_2 n \right\rceil + 1.
    \end{equation}}
\end{list}

These theorems form the mathematical basis for understanding the Non-Adjacent Form and its application in scalar multiplication.

\subsection{Theorem 7: Integer Representation in Base \( B \)}

\textbf{Theorem 7.}  
For a positive integer \( B \geq 2 \) and \( n \in \mathbb{N} \), there exist \( d \in \mathbb{N} \) and \( a_0, a_1, \ldots, a_{d-1} \in [0, B-1] \) such that:
\begin{equation}
    n = \sum_{i=0}^{d-1} a_i B^i.
\end{equation}

\textbf{Proof:}  
Let \( d := \left\lceil \log_B(n+1) \right\rceil \). Then \( d \in \mathbb{N} \), and for \( 0 \leq i \leq d-1 \), \( 0 \leq a_i < B \) holds.

\begin{equation}
    n = \sum_{i=0}^{d-1} a_i B^i = \sum_{i=0}^{d-1} \left( \left\lfloor \frac{n}{B^i} \right\rfloor \bmod B \right) B^i.
\end{equation}

This simplifies to:
\begin{equation}
    n = \sum_{i=0}^{d-1} \left( \left\lfloor \frac{n}{B^i} \right\rfloor - \left\lfloor \frac{n}{B^{i+1}} \right\rfloor B \right) B^i.
\end{equation}

Rearranging gives:
\begin{equation}
    n = \left\lfloor \frac{n}{B^0} \right\rfloor B^0 - \left\lfloor \frac{n}{B^d} \right\rfloor B^d.
\end{equation}

Since \( B^d \geq n+1 > n \), \( \left\lfloor \frac{n}{B^d} \right\rfloor = 0 \), hence:
\begin{equation}
    n = \left\lfloor \frac{n}{B^0} \right\rfloor B^0 = n.
\end{equation}

\subsection{Scalar Multiplication Using Theorem 7}

To compute scalar multiplication \( kP \), we represent \( k \) as:
\begin{equation}
    k = \sum_{i=0}^{d-1} a_i B^i, \quad 0 \leq a_i < B.
\end{equation}

Thus, scalar multiplication can be expressed as:
\begin{equation}
    kP = \sum_{i=0}^{d-1} a_i \left( B^i P \right).
\end{equation}

We precompute an array \( M \) defined as:
\begin{equation}
    M_{i,j} = j \left( B^i P \right), \quad 0 \leq i < d, \quad 0 \leq j < B.
\end{equation}

For \( j \geq 2 \), the array \( M \) satisfies:
\begin{equation}
    M_{i,j} = M_{i,j-1} + M_{i,1}.
\end{equation}

For \( j = 1 \):
\begin{equation}
    M_{i,1} = B^i P =
    \begin{cases} 
    M_{i-1, B}, & \text{if } i \geq 1, \\[6pt] 
    P, & \text{if } i = 0.
    \end{cases}
\end{equation}

For \( j = 0 \):
\begin{equation}
    M_{i,0} = O.
\end{equation}

The array \( M \) can be precomputed with a time complexity of \( \Theta(dB) \).

After precomputing \( M \), the scalar multiplication \( kP \) can be expressed as:
\begin{equation}
    kP = \sum_{i=0}^{d-1} a_i \left(B^i P\right) = \sum_{i=0}^{d-1} M_{i,a_i}.
\end{equation}

Thus, the time complexity for a single scalar multiplication is \( \Theta(d) \), and for \( Q \) scalar multiplications, it is \( \Theta(dQ) \). Including the precomputation time, the total complexity is:
\begin{equation}
    \Theta\left(d\left(B+Q\right)\right).
\end{equation}

\subsection{Minimization of Time Complexity}

Since the coefficient \( k \) could be as high as \( n-1 \), the parameters \( B \) and \( d \) must satisfy:

\begin{equation}
    B^d-1 \geq n-1.
\end{equation}

To minimize \( d(B + Q) \) with respect to \( B \), we choose \( B^* := \left \lceil \sqrt[d]{n} \right \rceil \). This leads to the overall time complexity:

\begin{equation}
    \Theta\left(d\left(B+Q\right)\right) = \Theta\left(d\left(\sqrt[d]{n}+Q\right)\right) = \Theta\left(d\left(\sqrt[d]{p}+Q\right)\right).
\end{equation}

The remaining variable is \( d \). To minimize:

\begin{equation}
    f(x) = x \left(p^{\frac{1}{x}} + Q\right), \quad x > 0,
\end{equation}

we find the derivative \( f'(x) = 0 \):

\begin{equation}
    x_0 = \frac{\ln p}{W\left(\frac{Q}{\mathrm{e}}\right) + 1},
\end{equation}

where \( W(\cdot) \) is the Lambert W function's principal branch. This indicates \( f(x) \) reaches its minimum at \( x = x_0 \).

Since \( f' \) is strictly increasing on \( (0, +\infty) \), \( f(x) \) is strictly decreasing on \( (0, x_0] \) and strictly increasing on \( [x_0, +\infty) \). Thus, \( f(x) \) attains its minimum at \( x = x_0 \).

\bibliographystyle{elsarticle-num}
\bibliography{reference}

\end{document}